\documentclass[a4paper,10pt,twocolumn]{article}

\usepackage[utf8]{inputenc}
\usepackage[OT1]{fontenc}
\usepackage{amsmath,graphicx,hyperref,ifthen,journals,microtype}
\usepackage[numbers]{natbib}
\usepackage[varg]{txfonts}
\usepackage[left=15mm,right=15mm,top=15mm,bottom=20mm]{geometry}
\newcommand{\latexonly}[1]{#1}
\newcommand{\wikionly}[1]{}

\title{Weak gravitational lensing}
\author{Matthias Bartelmann \and Matteo Maturi\\
  Universität Heidelberg, Zentrum für Astronomie, Institut für Theoretische Astrophysik}
\date{Invited and refereed contribution to
  \href{http://www.scholarpedia.org/article/Weak_gravitational_lensing}{Scholarpedia}}

\begin{document}

\maketitle

\begin{abstract}

According to the theory of general relativity, masses deflect light in a way similar to convex glass lenses. This gravitational lensing effect is astigmatic, giving rise to image distortions. These distortions allow to quantify cosmic structures statistically on a broad range of scales, and to map the spatial distribution of dark and visible matter. We summarise the theory of weak gravitational lensing and review applications to galaxies, galaxy clusters and larger-scale structures in the Universe.

\end{abstract}

\section{Brief historical outline of gravitational lensing}

\subsection{Newton and Soldner}

As early as in 1704, Sir Isaac Newton had surmised that light might be deflected by gravity. In his book ``Opticks: or, a treatise of the reflections, refractions, inflections and colours of light'', he collected a whole list of questions in an appendix, the first of which was: ``Do not Bodies act upon Light at a distance, and by their action bend its Rays; and is not this action (caeteris paribus) strongest at the least distance?'' \cite{NE04.1}

In Newtonian physics, however, this question cannot even be confidently addressed. Light can be considered as a stream of particles, the photons, which however have no rest mass. In the view of special relativity, photons acquire inertia by their motion, and this inertia contributes to their mass. However, if they could be at rest, no inertia and thus no mass would remain. But gravity should not be able to act on something without mass, or should it?

On the other hand, in Newtonian physics, the trajectory of a small test body around a large mass does not at all depend on the mass of the test body. As Galileo had noticed already, all bodies fall equally rapidly, provided that no forces besides gravity act upon them. If one imagines a light particle with even a tiny mass, its motion e.g. around the Sun would be calculable without problems. If this mass would be halved, the trajectory would remain the same. And what happened if its mass was progressively reduced until it finally completely disappeared?

The astronomer Johann Georg von Soldner was fully aware of this difficulty when he wrote his article ``On the deflection of a light ray from its straight motion by the attraction of a heavenly body which it passes closely'' for the Astronomical Almanac for the year 1804. Assuming that a mass could (and should) be assigned to the (then hypothetical) particles of light, he calculated by what amount such light rays would be deflected that, as seen from the Earth, would just graze the Solar rim. Almost apologetically, he wrote: ``Hopefully nobody will find it questionable that I treat a light ray perfectly as a heavy body. For that light rays have all the absolute properties of matter is seen by the phenomenon of aberration, which is possible only by light rays being indeed material. – And besides, nothing can be conceived that exists and acts on our senses without having the properties of matter.'' \cite[][originally in German, our translation]{SO04.1}

\subsection{Einstein's two deflection angles}

In Einstein's perception of the essence of gravity, however, this problem does not even exist. Einstein's theory of general relativity no longer interprets gravity as a force in the Newtonian sense, but explains the effects of gravity on the motion of bodies by the curvature of space-time caused by the presence of matter or energy.

In the framework of an unfinished version of the theory of general relativity, Einstein calculated the deflection of light at the rim of the Sun for the first time in 1911. He found: ``A light ray passing the Sun would accordingly suffer a deflection of the magnitude $4\cdot10^{-6} = 0.83$ arc seconds.'' \cite{EI11.1} This was exactly the same value that Johann Georg von Soldner had found by his calculation within Newtonian physics. The second time, in 1916, Einstein arrived at the conclusion: ``A light ray passing the Sun thus experiences a deflection of 1.7 arc seconds.'' \cite[][originally in German, our translation]{EI16.1} His value had plainly doubled, and for a good reason: Only within the final theory of general relativity did Einstein find that he needed to account for temporal as well as spatial curvature of space-time by gravity, and this caused precisely twice the amount of light deflection to occur.

This value of 1.7 arc seconds, twice as large as expected from the Newtonian calculation, was confirmed shortly thereafter by observations. During a total Solar eclipse on May 29, 1919, two British expeditions succeeded in measuring the angle by which such stars appeared pushed away from the Sun which happened to be close to the Sun at the moment of the eclipse and became momentarily visible while the Sun was obscured. In November 1919, the authors Dyson, Eddington and Davidson reported: ``Thus the results of the expeditions to Sobral and Principe can leave little doubt that a deflection of light takes place in the neighbourhood of the Sun and that it is of the amount demanded by Einstein's general theory of relativity, as attributable to the sun's gravitational field.'' \cite{ED19.1}

On October 9, 1919, Einstein himself had reported in a brief note: ``According to a telegram addressed to the signatory by Prof.\ Lorentz, the English expedition under Eddington, sent out to observe the Solar eclipse of May 29, has observed the deflection of light at the rim of the Sun required by the general theory of relativity. The value provisionally deduced so far falls between 0.9 and 1.8 arc seconds. The theory demands 1.7 arc seconds.'' \cite[][originally in German, our translation]{EI19.1}

\subsection{Mandl and Zwicky}

Even though, for Einstein himself, the deflection of light at the rim of the Sun marked a confirmation of his theory of general relativity, he was sure that this effect would hardly ever gain astrophysical relevance, not to speak of an even more practical importance. When the Czech engineer Rudi W. Mandl visited Einstein in Princeton in 1936 and asked him to calculate the gravitational lensing effect of a star on a more distant star, Einstein finally consented to publish a note on this phenomenon \cite{EI36.1}, which he however believed to be unobservable. Indirectly also stimulated by Mandl, the Swiss-American astronomer Fritz Zwicky, however, voiced the idea in 1937 that entire clusters of galaxies could act as gravitational lenses \cite{ZW37.1}. Only four years earlier, Zwicky had found the first indication of dark matter by observations of the Coma galaxy cluster \cite{ZW33.1}. Nonetheless, it took until 1979 for the first gravitational lens to be found \cite{WA79.1}. This lens is a galaxy embedded into a group of galaxies whose gravitational-lensing effect turns a bright, approximately point-like, very distant object, a so-called quasi-stellar object or QSO, into two images.

\section{Deflection angle and lensing potential}

\subsection{Index of refraction and Fermat's principle}

There are multiple ways at gravitational lensing which agree in the limit which is commonly applied. In by far the most astrophysical applications, the Newtonian gravitational potential $\Phi$ is small, $|\Phi|/c^2 \ll 1$, and the lensing mass distribution moves slowly with respect to the cosmological rest frame. In a galaxy cluster, for example, $|\Phi|/c^2 \lesssim 10^{-5}$, and cosmic structures have typical peculiar velocities $v \lesssim 600\,\mathrm{km\,s^{-1}} \ll c$. The standard approach to gravitational lensing has been laid out in several reviews, lecture notes and a text book \latexonly{\cite{SC92.1, NA99.1, ME99.2, BA01.2, RE03.1, ME06.1, BA10.2}}\wikionly{\cite{SC92.1, BA01.2, BA10.2}}.

Under such conditions, gravitational lensing can be described by a small perturbation of the locally Minkowskian space-time of an observer co-moving with the gravitational lens. The Minkowski metric of special relativity, expressed by its line element
\begin{equation}
  \mathrm{d}s^2 = -c^2\mathrm{d}t^2+\mathrm{d}\vec x^{\,2}\;,
\label{eq:1}
\end{equation}
is perturbed by the dimension-less Newtonian gravitational potential $\Phi/c^2$ as
\begin{equation}
  \mathrm{d}s^2 = -c^2\left(1+\frac{2\Phi}{c^2}\right)\mathrm{d}t^2+
  \left(1-\frac{2\Phi}{c^2}\right)\mathrm{d}\vec x^{\,2}\;.
\label{eq:2}
\end{equation}
With the propagation condition for light, $\mathrm{d}s = 0$, this expression can be rearranged to find the effective light speed in a weak gravitational field,
\begin{equation}
  c' = \left\vert\frac{\mathrm{d}\vec x}{\mathrm{d}t}\,\right\vert =
  c\left(1+\frac{2\Phi}{c^2}\right)\;,
\label{eq:3}
\end{equation}
where $\Phi/c^2\ll 1$ was used in a first-order Taylor expansion. Introducing the index of refraction $n$ by the conventional definition $c' = c/n$, we see that a weak gravitational field has the \emph{effective index of refraction}
\begin{equation}
  n = \frac{c}{c'} = 1-\frac{2\Phi}{c^2}\;.
\label{eq:4}
\end{equation}
Since the gravitational potential is negative if conventionally normalised such as to vanish at infinity, this index of refraction is larger than unity.

We can now apply \emph{Fermat's principle}, which asserts that a light ray follows that path between two fixed points $A$ and $B$ along which its optical path $\tau$ is extremal,
\begin{equation}
  \delta\tau = \delta\int_A^B\frac{c}{n}\,\mathrm{d}t = 0\;.
\label{eq:5}
\end{equation}
The variation of $\tau$ with respect to the light path leads directly to the deflection angle
\begin{equation}
  \hat{\vec\alpha} = -\frac{2}{c^2}\int\vec\nabla_\perp\Phi\,\mathrm{d}\lambda\;,
\label{eq:6}
\end{equation}
which is the gradient of the dimension-less Newtonian potential perpendicular to the light ray, integrated along the light ray and multiplied by two. This factor of two comes from the fact that the perturbed Minkowski metric has equal perturbations in both its temporal and spatial components, as mentioned above in the context of Einstein's two calculations.

\subsection{The Born approximation}

The integral over the actual light path is complicated to carry out. However, since typical deflection angles are on the order of arc seconds or smaller, the integration path can be approximated by a straight line, just as in the \emph{Born approximation} familiar from quantum mechanics. Then, assuming a point mass $M$ at the origin of a coordinate system and a light ray propagating parallel to the $z$ axis and passing the point mass at an impact parameter $b$, the deflection angle according to (\ref{eq:6}) is
\begin{equation}
  \hat\alpha = -\frac{2}{c^2}\frac{\partial}{\partial b}
  \int_{-\infty}^\infty\mathrm{d}z\,\frac{GM}{\sqrt{b^2+z^2}} =
  \frac{4GM}{bc^2} = \frac{2R_\mathrm{S}}{b}\;,
\label{eq:7}
\end{equation}
where $R_\mathrm{S}=2GM/c^2$ is the Schwarzschild radius of the lensing point mass. For the Sun, with its mass of $M = 2\cdot10^{33}\,\mathrm{g}$, the Schwarzschild radius is $R_{\mathrm{S}\odot} \approx 3\,\mathrm{km}$, hence the deflection angle at the Solar radius $R_\odot = 7\cdot10^5\,\mathrm{km}$ is
\begin{equation}
  \hat\alpha_\odot \approx \frac{6}{7\cdot10^5} \approx 8.6\cdot10^{-6}
  \approx 1.7''\;.
\label{eq:8}
\end{equation}
This is Einstein's famous result, tested and verified by Dyson, Eddington and Davidson in 1919.

\subsection{The lensing potential}

If the lensing mass distribution is thin compared to the overall extent of the lens system, the light path from the observer to the source can be approximated by straight lines from the observer to the lens and from there on to the source which enclose the deflection angle. This is called the \emph{thin-lens approximation}. It is appropriate for the description of isolated lenses such as galaxy clusters, but inadequate for extended lenses such as the large-scale structures of the Universe. We shall proceed with the thin-lens approximation for now and later generalise the results to extended mass distributions.

\begin{figure}[ht]
  \centerline{\includegraphics[width=\hsize]{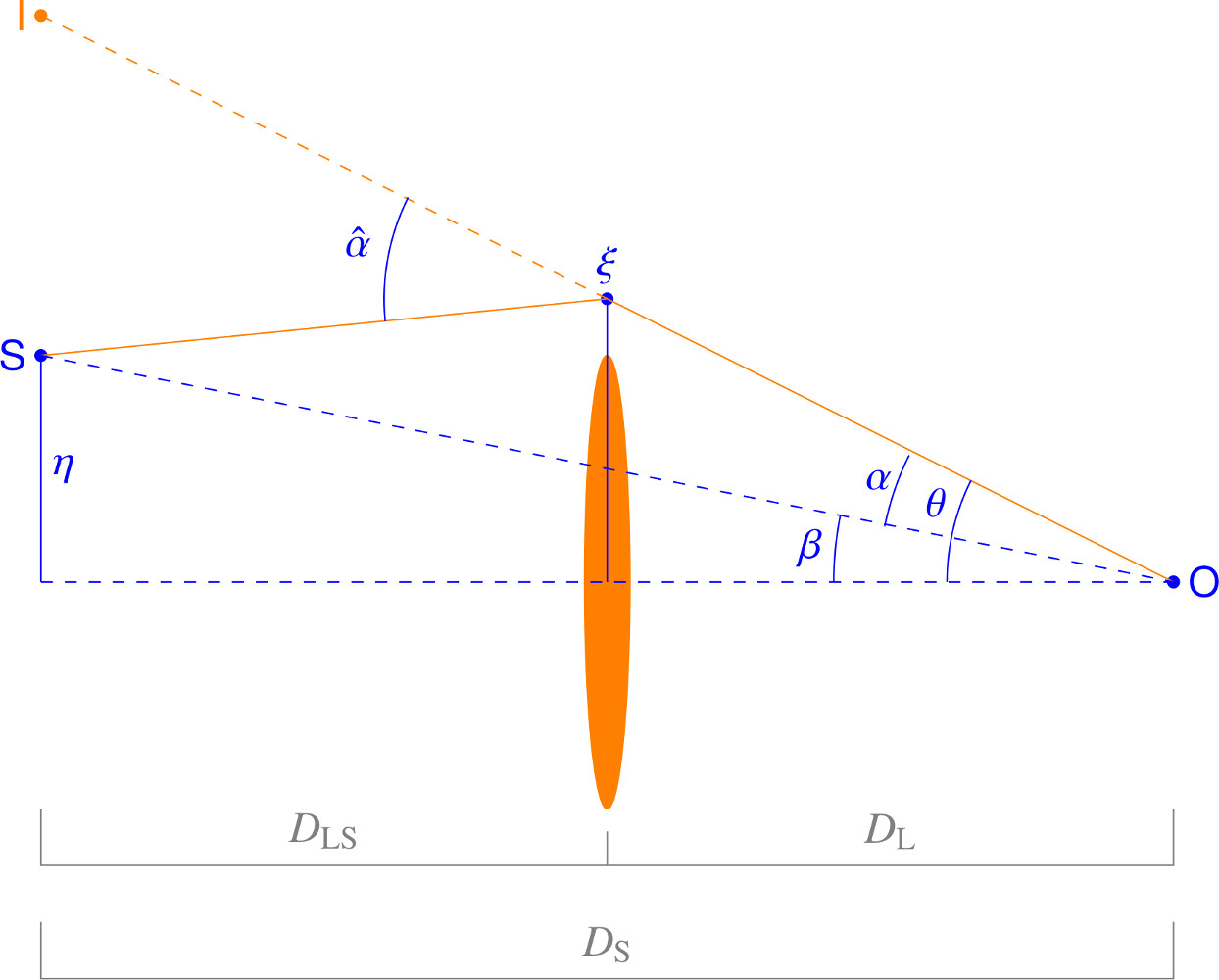}}
\caption{Sketch of a gravitational-lens system (adapted from \cite{NA99.1}): The optical axis runs from the observer O through the centre of the lens. The angle between the source S and the optical axis is $\beta$, the angle between the image I and the optical axis is $\theta$. The light ray towards the image is bent by the deflection angle $\hat\alpha$, measured at the lens. The reduced deflection angle $\alpha$ is measured at the observer.}
\label{fig:01}
\end{figure}

Tracing a light ray back from the observer to the source, the lens or ray-tracing equation can then simply be found from the intercepts of the light ray relative to the (arbitrary) optical axis at the distances of the lens and the source from the observer, $D_\mathrm{L}$ and $D_\mathrm{S}$, respectively. Let the angles between the optical axis and the image, and between the optical axis and the source on the observer's sky be $\vec\theta$ and $\vec\beta$, respectively. These angles are vectors because they have a direction as well as a magnitude. Then,
\begin{equation}
  D_\mathrm{S}\vec\beta = D_\mathrm{S}\vec\theta-D_\mathrm{LS}\hat{\vec\alpha}\;,
\label{eq:9}
\end{equation}
with $D_\mathrm{LS}$ being the distance between the lens and the source (cf.\ Fig.\ \ref{fig:01}). This equation may look trivial, but it is not. It only holds because even in the curved space-time of the Universe, distances can be introduced in such a way that the intercept theorem familiar from Euclidean geometry continues to hold. These are the \emph{angular-diameter distances}. Dividing by $D_\mathrm{S}$ and introducing the \emph{reduced deflection angle}
\begin{equation}
  \vec\alpha := \frac{D_\mathrm{LS}}{D_\mathrm{S}}\,\hat{\vec\alpha}
\label{eq:10}
\end{equation}
leads to the lens or ray-tracing equation in its simplest form,
\begin{equation}
  \vec\beta = \vec\theta-\vec\alpha\;.
\label{eq:11}
\end{equation}
With Eq. (\ref{eq:6}), the reduced deflection angle can be written as a gradient,
\begin{equation}
  \vec\alpha = \vec\nabla_\perp\left[
    \frac{2}{c^2}\frac{D_\mathrm{LS}}{D_\mathrm{S}}\int\Phi\,\mathrm{d}z
  \right]\;.
\label{eq:12}
\end{equation}

It is convenient to introduce the quantities describing gravitational lensing, such as the reduced deflection angle or the lensing potential, as functions on the sky, i.e.\ as functions of the angular position $\vec\theta$ on the celestial sphere. Then, gradients need to be taken with respect to angles rather than perpendicular distances. The perpendicular gradient $\vec\nabla_\perp$ occurring in (\ref{eq:6}) and (\ref{eq:12}) is then replaced by the gradient $\vec\nabla_\theta$ with respect to the angle $\vec\theta$ according to
\begin{equation}
  \vec\nabla_\perp = D_\mathrm{L}^{-1}\vec\nabla_\theta\;.
\label{eq:13}
\end{equation}
The distance $D_\mathrm{L}$ from the observer to the lens appears here because, in the small-angle approximation, the perpendicular separation from the line-of-sight is $D_\mathrm{L}\vec\theta$. Thus, (\ref{eq:12}) turns into
\begin{equation}
  \vec\alpha = \vec\nabla_\theta\psi\quad\mbox{with}\quad
  \psi := \frac{2}{c^2}
  \frac{D_\mathrm{LS}}{D_\mathrm{L}D_\mathrm{S}}\int\Phi\,\mathrm{d}z\;.
\label{eq:14}
\end{equation}
The quantity $\psi$ defined here is called the \emph{lensing potential}. In by far the most situations of astrophysical interest, the lensing potential incorporates all imaging properties of a gravitational lens.

For a point-mass lens, we have seen in (\ref{eq:7}) that the deflection angle $\hat\alpha$ is proportional to the inverse impact parameter, $b^{-1}$. Converting $b$ to the angle $\theta$ by $b = D_\mathrm{L}\theta$ and the deflection angle $\hat\alpha$ to the reduced
deflection angle $\alpha$, we find
\begin{equation}
  \alpha = \frac{\partial\psi}{\partial\theta} \quad\mbox{with}\quad
  \psi = \frac{4GM}{c^2}
  \frac{D_\mathrm{LS}}{D_\mathrm{L}D_\mathrm{S}}
  \ln\left\vert\theta\right\vert\;,
\label{eq:15}
\end{equation}
showing that the lensing potential of a point-mass lens at the coordinate origin is proportional to the logarithm of the angular radius.

\section{Magnification and distortion}

\subsection{The convergence}

Taking the divergence of $\vec\alpha$ leads to the Laplacian of $\psi$,
\begin{equation}
  \vec\nabla_\theta\cdot\vec\alpha = \vec\nabla_\theta^2\psi =
  \frac{2}{c^2}
  \frac{D_\mathrm{L}D_\mathrm{LS}}{D_\mathrm{S}}
  \int\vec\nabla_\perp^2\Phi\,\mathrm{d}z\;,
\label{eq:16}
\end{equation}
where we have replaced the Laplacian $\vec\nabla_\theta^2$ with respect to the angle $\vec\theta$ by the perpendicular Laplacian $\vec\nabla_\perp^2$ with respect to physical coordinates. If we could replace the perpendicular by the complete Laplacian,
\begin{equation}
  \vec\nabla^2 = \vec\nabla_\perp^2+\frac{\partial^2}{\partial z^2}\;,
\label{eq:17}
\end{equation}
we could insert \emph{Poisson's equation}
\begin{equation}
  \vec\nabla^2\Phi = 4\pi G\rho
\label{eq:18}
\end{equation}
into (\ref{eq:16}). We can indeed do so if the gradient of the potential $\Phi$ along the line-of-sight, taken at the beginning and at the end of the line-of-sight, can be neglected,
\begin{equation}
  \int\frac{\partial^2\Phi}{\partial z^2}\,\mathrm{d}z =
  \left.\frac{\partial\Phi}{\partial z}\right\vert_\mathrm{end~points} = 0\;.
\label{eq:19}
\end{equation}
This is indeed excellently satisfied in all situations where the extent of the lensing mass distribution is small compared to the cosmological distances $D_\mathrm{L}$, $D_\mathrm{LS}$ and $D_\mathrm{S}$ characterising the geometry of the lens system.

We thus substitute the complete, three-dimensional Laplacian for the two-dimensional, perpendicular Laplacian in (\ref{eq:16}) and use Poisson's equation to write
\begin{equation}
  \vec\nabla_\theta^2\psi =
  \frac{8\pi G}{c^2}\frac{D_\mathrm{L}D_\mathrm{LS}}{D_\mathrm{S}}\,\Sigma\;,
\label{eq:20}
\end{equation}
where the \emph{surface mass density}
\begin{equation}
  \Sigma := \int\rho\,\mathrm{d}z
\label{eq:21}
\end{equation}
is defined as the line-of-sight projection of the three-dimensional mass density $\rho$. The prefactor
\begin{equation}
  \frac{4\pi G}{c^2}\frac{D_\mathrm{L}D_\mathrm{LS}}{D_\mathrm{S}} =:
  \Sigma_\mathrm{cr}^{-1}
\label{eq:22}
\end{equation}
has the dimension of an inverse surface-mass density, $\mathrm{cm^2\,g^{-1}}$. The quantity $\Sigma_\mathrm{cr}$ is called \emph{critical surface mass density}. With these definitions,
\begin{equation}
  \vec\nabla_\theta^2\psi = 2\,\frac{\Sigma}{\Sigma_\mathrm{cr}} =: 2\kappa\;,
\label{eq:23}
\end{equation}
where we introduced the dimension-less surface-mass density or \emph{convergence} $\kappa$. From here on, we shall drop the subscript $\theta$ on the gradient, understanding that $\vec\nabla$ is the gradient $\vec\nabla_\theta$ with respect to $\vec\theta$.

\subsection{Geometrical sensitivity of gravitational lensing}

The Poisson equation (\ref{eq:23}) shows that the source of the lensing potential $\psi$ is twice the dimension-less convergence or surface-mass density $\kappa$, i.e.\ the surface-mass density $\Sigma$ of the lens divided by its critical surface mass density $\Sigma_\mathrm{cr}$. A mass distribution with a fixed surface-mass density $\Sigma$ can thus be a more or less efficient gravitational lens, depending on the overall extent of the lens system composed of observer, source, and lens, and depending on where the lens is located along the line-of-sight. Lensing is most efficient where the critical surface-mass density $\Sigma_\mathrm{cr}$ is minimal, or where the expression in (\ref{eq:22}) is maximal. In Euclidean space, this would be half-way between the observer and the source. In the curved space-time of our Universe, the location of maximal lensing sensitivity is somewhat closer in redshift to the observer. The geometrical sensitivity of gravitational lensing is one of the main characteristics turning lensing into a powerful tool for cosmology.

\subsection{Linearised lens mapping and Jacobi matrix}

In terms of the lensing potential $\psi$, the lens equation (\ref{eq:11}) is
\begin{equation}
  \vec\beta = \vec\theta-\vec\nabla\psi\;.
\label{eq:24}
\end{equation}
Imagine now a source substantially smaller than any typical scale of variation in the deflection angle. Let $\delta\vec\beta$ the angular separation of a point on an outer source contour from the centre of the source. Then, the corresponding angular distance of the image point can be approximated by a first-order Taylor expansion of the lens equation (\ref{eq:24}),
\begin{equation}
  \delta\vec\beta \approx \mathcal{A}\,\delta\vec\theta\;,
\label{eq:25}
\end{equation}
where $\mathcal{A}$ is the \emph{Jacobian matrix} of the lens mapping. It has the components
\begin{equation}
  \mathcal{A}_{ij} = \frac{\partial\beta_i}{\partial\theta_j} =
  \delta_{ij}-\psi_{ij}\;,
\label{eq:26}
\end{equation}
where the potential derivatives are to be taken at the centre of the lensed image. Here, we have introduced the common short-hand notation
\begin{equation}
  \psi_{ij} := \frac{\partial^2\psi}{\partial\theta_i\partial\theta_j}
\label{eq:27}
\end{equation}
for the second partial derivatives of $\psi$.

Equations (\ref{eq:25}) and (\ref{eq:26}) are interesting expressions. First, (\ref{eq:25}) states that the Jacobi matrix $\mathcal{A}$ maps small distances $\delta\vec\theta$ in an image back to small distances $\delta\vec\beta$ in the source. This is the foundation of by far the most applications of weak gravitational lensing in astrophysics and cosmology. Equation (\ref{eq:26}) shows that, in the absence of the lensing potential, the lens mapping is simply the identity. In the presence of a lens, the local properties of the lens mapping are determined by the curvature of the lensing potential $\psi$, expressed by the matrix of second derivatives of $\psi$, or the Hessian matrix of the potential. Second derivatives of gravitational potentials are tidal forces. Locally, deformations by the lens mapping are thus determined by the gravitational tidal forces caused by the lens.

\subsection{Shear and magnification}

For the physical interpretation of the Jacobi matrix $\mathcal{A}$, it is convenient and instructive to split $\mathcal{A}$ into an isotropic and an anisotropic, trace-free part by taking the trace,
\begin{equation}
  \mathrm{tr}\mathcal{A} = 2-\vec\nabla^2\psi = 2(1-\kappa)\;,
\label{eq:28}
\end{equation}
and subtracting it from $\mathcal{A}$ by means of the unit matrix $\mathcal{I}$ to obtain the shear matrix
\begin{equation}
  \Gamma := -\left(
    \mathcal{A}-\frac{1}{2}\left(\mathrm{tr}\mathcal{A}\right)\mathcal{I}
  \right)
\label{eq:29}
\end{equation}
which has the components
\begin{align}
  \Gamma_{11} &=: \gamma_1 =
  \frac{1}{2}\left(\psi_{11}-\psi_{22}\right)\;,\quad
  \Gamma_{22} = -\gamma_1\;,\nonumber\\
  \Gamma_{12} &= \Gamma_{21} =: \gamma_2 = \psi_{12}\;.
\label{eq:30}
\end{align}
These manipulations leave the Jacobi matrix in the form
\begin{equation}
  \mathcal{A} = (1-\kappa)\mathcal{I}-\Gamma =
  \left(\begin{array}{cc}
    1-\kappa-\gamma_1 & -\gamma_2\\-\gamma_2 & 1-\kappa+\gamma_1
  \end{array}\right)\;.
\label{eq:31}
\end{equation}

The linearised lens equation (\ref{eq:25}) tells us the inverse of what we typically want to know from weak gravitational lensing. Since we observe images but cannot access the sources, we need to infer the separation $\delta\vec\theta$ of image points from the image centre. If the Jacobi matrix $\mathcal{A}$ has a non-vanishing determinant, it can be inverted, allowing us to write
\begin{equation}
  \delta\vec\theta = \mathcal{A}^{-1}\delta\vec\beta\;.
\label{eq:32}
\end{equation}
The inverse Jacobi matrix determines how sources are mapped on images. In weak gravitational lensing, the Jacobi determinant
\begin{equation}
  \det\mathcal{A} = (1-\kappa)^2-\gamma^2
\label{eq:33}
\end{equation}
with
\begin{equation}
  \gamma^2 := \gamma_1^2+\gamma_2^2
\label{eq:34}
\end{equation}
is near unity because the absolute values of the convergence $\kappa$ and the shear $\gamma$ are both small compared to unity. Points in the lens where $\det\mathcal{A} = 0$ are called critical points. They play no role in weak gravitational lensing but are centrally important for strong lensing.

Thus, we can always assume in weak gravitational lensing that the linear lens mapping is invertible. The inverse of the Jacobi matrix is
\begin{equation}
  \mathcal{A}^{-1} = \frac{1}{\det\mathcal{A}}
  \left(\begin{array}{cc}
    1-\kappa+\gamma_1 & \gamma_2\\ \gamma_2 & 1-\kappa-\gamma_1
  \end{array}\right)\;.
\label{eq:35}
\end{equation}
The prefactor in this expression indicates that the solid angle spanned by the image is changed compared to the solid angle covered by the source by the \emph{magnification factor}
\begin{equation}
  \mu = \frac{1}{\det\mathcal{A}} = \frac{1}{(1-\kappa)^2-\gamma^2} \approx
  1+2\kappa\;,
\label{eq:36}
\end{equation}
where the final approximation is a first-order Taylor expansion. Thus, in weak lensing, the magnification of an image is essentially (i.e.\ to first Taylor order) determined by the convergence $\kappa$, not by the shear.

\subsection{Image distortion}

\begin{figure}[ht]
  \centerline{\includegraphics[angle=90, width=\hsize]{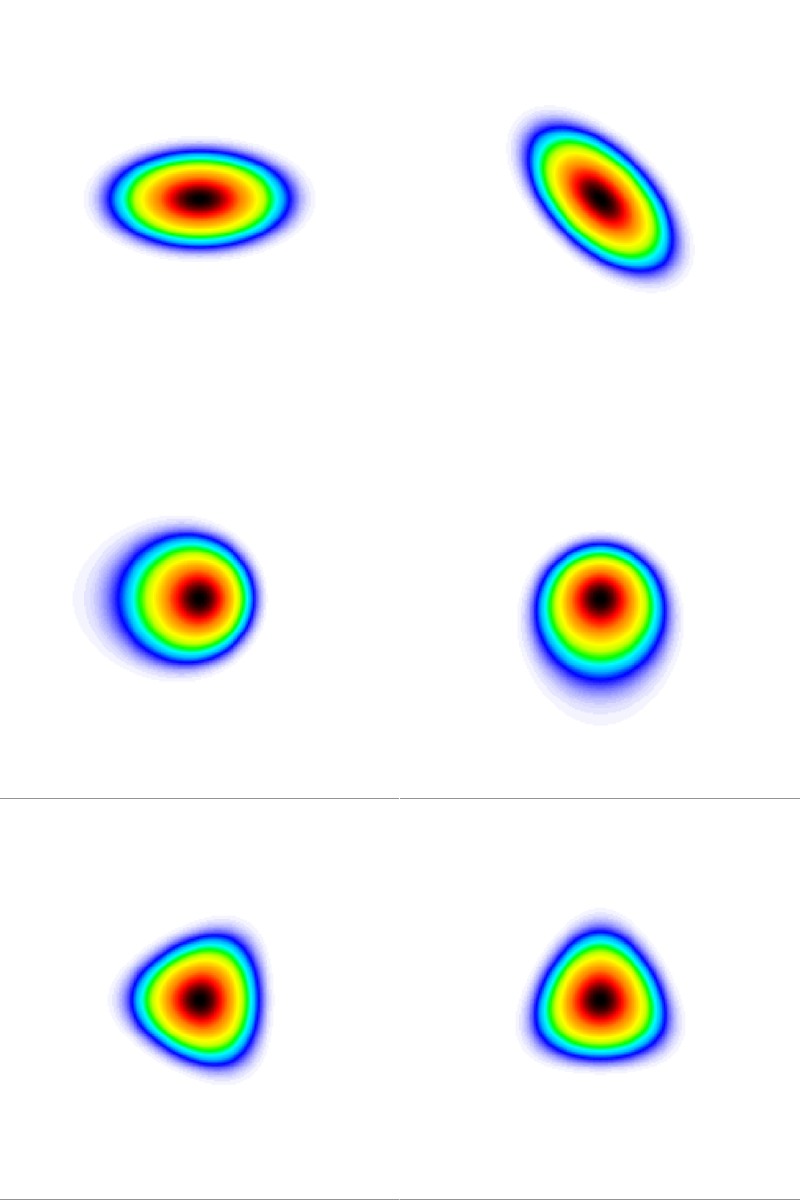}}
\caption{Illustration of gravitational-lensing effects on small sources: The left column shows the effects of both components of the shear on a circular source. The middle and right columns illustrate effects of higher order, quantified by the so-called flexion components $\mathcal{F}$ and $\mathcal{G}$, which are linear combinations of third potential derivatives. $\mathcal{F}$ flexion leads to a centroid shift while $\mathcal{G}$ flexion causes a triangular deformation.}
\label{fig:02}
\end{figure}

The eigenvalues of the inverse Jacobi matrix $\mathcal{A}^{-1}$ are
\begin{equation}
  \lambda_\pm = \frac{1-\kappa\pm\gamma}{\det\mathcal{A}} =
  \frac{1}{1-\kappa\mp\gamma}\;.
\label{eq:37}
\end{equation}
A hypothetical circular source is deformed by weak gravitational lensing to become an ellipse whose semi-major and semi-minor axes, called $a$ and $b$, respectively, are proportional to the eigenvalues $\lambda_\pm$. By its common definition, the \emph{ellipticity} $\varepsilon$ of such an image is
\begin{equation}
  \varepsilon := \frac{a-b}{a+b} =
  \frac{\lambda_+-\lambda_-}{\lambda_++\lambda_-} =
  \frac{\gamma}{1-\kappa}\;.
\label{eq:38}
\end{equation}
Since the source size prior to lensing is typically inaccessible to observation, the magnification of an image relative to the source cannot directly be measured. The ellipticity $\varepsilon$ and the orientation of elliptically distorted images are thus the only information supplied by weak lensing in the vast majority of cases. Equation (\ref{eq:38}) shows that the ellipticity is determined by the so-called \emph{reduced shear}
\begin{equation}
  g = \frac{\gamma}{1-\kappa}
\label{eq:39}
\end{equation}
rather than by the shear itself. If $\kappa\ll 1$ which can most often be assumed in cases of weak lensing, $\varepsilon=g\approx\gamma$. The effects of both shear components on a circular source are illustrated in Fig.\ \ref{fig:02} together with higher-order distortions.

\section{Ellipticity measurement}

\subsection{Local averages and angular resolution}

The sources typically used in observations of weak gravitational lensing are of course not circular, but intrinsically elliptical. In lowest-order approximation, the intrinsic source ellipticity $\varepsilon_\mathrm{S}$ and the ellipticity caused by lensing add. Then,
\begin{equation}
  \varepsilon \approx g+\varepsilon_\mathrm{S}\;.
\label{eq:40}
\end{equation}
One of the most essential assumptions in the interpretation of weak gravitational lensing is that the intrinsic ellipticities approach zero when averaged over sufficiently large samples, $\langle\varepsilon_\mathrm{S}\rangle \approx 0$.

Fortunately, the sky is studded with so many faint, distant galaxies that averages over many of them can be taken on angular scales small compared to the typical scales which the shear and the convergence of the lens vary on. On deep images taken with ground-based telescopes (at an $r$-band magnitude of $\sim25$), $n\approx 10$ galaxies are found per square arc minute, and $n\approx 30-50$ is reached on images taken in space. The standard deviation of the intrinsic ellipticity is measured to be $\sigma_\varepsilon \approx 0.2$ per source. Averaging over $N$ faint galaxy images, the scatter of the intrinsic ellipticity is reduced to
\begin{equation}
  \Delta\left\langle\varepsilon_\mathrm{S}\right\rangle \approx
  \frac{\sigma_\varepsilon}{\sqrt{N}}\;,
\label{eq:41}
\end{equation}
and the angular resolution $\Delta\theta$ of this measurement is limited by
\begin{equation}
  N = \pi\Delta\theta^2n \quad\mbox{or}\quad
  \Delta\theta = \left(\frac{N}{\pi n}\right)^{1/2}\;.
\label{eq:42}
\end{equation}
For $N=10$, we find $\Delta\langle\varepsilon_\mathrm{S}\rangle\approx0.06$ and $\Delta\theta\approx1'$, assuming $n\approx10$ source galaxies per square arc minute. It is of paramount importance for weak gravitational lensing to verify that intrinsic source ellipticities do indeed average to zero, or otherwise to quantify the degree to which they do not.

\subsection{Measurement process}
\label{sec:4.2}

There are essentially two general approaches to the measurement of image ellipticities. Both of them build upon the measured surface brightness $I(\vec\theta\,)$ of an image. In the forward modelling approach, models for the surface brightness of elliptical sources are fit to the image, allowing to read off the model ellipticity once the best fit has been found. In the other, model-free approach, the quadrupole moments $Q_{ij}$ of the surface brightness are measured,
\begin{equation}
  Q_{ij} = \int\mathrm{d}^2\theta\,I(\vec\theta\,)\theta_i\theta_j\;,\quad i,j = 1,2\;.
\label{eq:43}
\end{equation}
From them, the two components of the ellipticity are
\begin{equation}
  \varepsilon_1 = \frac{Q_{11}-Q_{22}}{2N_Q}\;,\quad
  \varepsilon_2 = \frac{Q_{12}}{N_Q}\;,\quad
  N_Q := \frac{1}{2}\mathrm{tr}Q+\sqrt{\det Q}\;.
\label{eq:44}
\end{equation}

Both methods have their advantages and disadvantages. The forward-modelling approach needs to assume a surface-brightness distribution which may or may not reflect the image properties well. Biases in the inferred ellipticities are possible and likely if the model is not well adapted to the image. The measurement of the surface-brightness moments is hampered by the fact that the formally infinite integral in (\ref{eq:43}) will pick up intensity fluctuations surrounding the typically faint and small image. To control this noise contribution, a weight function needs to be introduced in (\ref{eq:43}), effectively limiting the domain of the integral. The inevitable effects of the weight function on the ellipticity measurement need to be corrected afterwards. Alternatively, denoising procedures can be applied to the images.

On the whole, measuring the ellipticities of distant source galaxies accurately is very difficult. Distortions caused by weak lensing are on the order of up to $\sim10$ per cent, i.e.\ the difference of the semi-axes of typical images is only a few per cent of their sum. Ellipticities thus have to be measured either by model fitting or by measuring the quadrupole moments of the surface-brightness distribution from faint, small, pixellised images distorted by weak lensing at the per cent level.

Substantial complications arise because these faint galaxy images are irregular and distorted prior to lensing. Moreover, they are being observed through optical systems with their own small, but numerous and subtle imperfections and physical limitations. Even if small, these imperfections typically imprint noise and systematics on the measured ellipticities which also have to be carefully corrected. In particular, the point-spread function of the telescope optics needs to be carefully determined in order to quantify any distortions caused by the optical system itself. Usually, the point-spread function is estimated from the images of stars in the observed field.

In the course of these developments, algorithms were described and implemented for measuring the weak shear signal from large data fields. The Shear Testing Programme (STEP) was launched to test and improve the accuracy of shear measurements from weakly distorted images of distant galaxies in the presence of several perturbing effects \latexonly{\cite{HE06.1, MA07.1}}.

On the whole, ellipticity measurements must still be considered an art which is under ongoing development \latexonly{\cite{BR09.2, ME09.3, ME11.1, ME12.1, RE12.1, VI12.1, MA13.1, MI13.1}}. For the purposes of this review, we cannot go into any detail of this complicated and demanding measurement process. Suffice it to say that it belongs to the most fascinating developments in extragalactic astrophysics and cosmology since the turn of the century that highly reliable and significant measurements of weak gravitational lensing by galaxies, galaxy clusters as well as by large-scale structures have routinely become possible.

\section{Extended lenses}

\subsection{Generalisation of the lensing potential}

So far, we have assumed that the lens is geometrically small compared to the overall scale of the lens system. Since this is clearly not appropriate for gravitational lensing by the large-scale structures in our Universe, we need to generalise the results derived so far to the case of extended lenses. To achieve this, it suffices to pull the distance factors in the definition of the lensing potential in (\ref{eq:14}) under the line-of-sight integral. Before we can meaningfully do so, however, we need to clarify the concept of a distance in cosmology.

In the curved and expanding space-time of our Universe, distances are no longer uniquely defined. Depending on the measurement procedure intended, distances typically turn out to be vastly different. For gravitational lensing, the appropriate distance measure is defined such that the ratio between the physical size of a small object and its angular extent equals its distance, which is the relation familiar from static Euclidean space. The distance defined this way is called \emph{angular-diameter distance}.

Since our Universe is expanding, spatial separations between any two points grow in time in a way quantified by the so-called scale factor $a$. Commonly, the scale factor is normalised to unity at the present cosmic epoch. It is then convenient to introduce so-called \emph{comoving distances}, which are distances defined on a spatial hypersurface constructed at the present time. Measurements have shown that, even though our Universe has a finite space-time curvature, its spatial curvature cannot be distinguished from zero within the measurement uncertainty. In other words, our Universe turns out to be well described as spatially flat.

Let now $\chi$ be the comoving angular-diameter distance in a spatially flat space-time, then the lensing potential of an extended lens acting on a source at distance $\chi_\mathrm{S}$ is
\begin{equation}
  \psi(\vec\theta\,) = \frac{2}{c^2}\int_0^{\chi_\mathrm{S}}\mathrm{d}\chi\,
  \frac{\chi_\mathrm{S}-\chi}{\chi_\mathrm{S}\chi}\,
  \Phi(\chi\vec\theta,\chi)\;.
\label{eq:45}
\end{equation}
Note the similarity to the expression for the lensing potential of a thin lens in (\ref{eq:14}): the distance prefactor is replaced according to
\begin{equation}
  \frac{D_\mathrm{LS}}{D_\mathrm{L}D_\mathrm{S}} \to
  \frac{\chi_\mathrm{S}-\chi}{\chi_\mathrm{S}\chi}
\label{eq:46}
\end{equation}
and pulled under the integral, the line-of-sight integration is being performed over $\mathrm{d}\chi$, and the Newtonian potential is taken at the position $\chi\vec\theta$ perpendicular and $\chi$ parallel to the line-of-sight.

\subsection{Deflection angle, convergence and shear}

The remaining quantities, in particular the reduced deflection angle $\vec\alpha$, the convergence $\kappa$ and the shear $\gamma$, can now be derived from $\psi$ in the usual manner,
\begin{equation}
  \vec\alpha = \vec\nabla\psi\;,\quad
  \kappa = \frac{1}{2}\vec\nabla^2\psi\;,\quad
  \gamma_1 = \frac{1}{2}\left(\psi_{11}-\psi_{22}\right)\;,\quad
  \gamma_2 = \psi_{12}\;,
\label{eq:47}
\end{equation}
where all derivatives are to be taken with respect to the angular position $\vec\theta$, as introduced before. In particular, the convergence to be assigned to an extended lens is
\begin{equation}
  \kappa = \frac{4\pi G}{c^2}\int_0^{\chi_\mathrm{S}}\mathrm{d}\chi\,
  \frac{\chi(\chi_\mathrm{S}-\chi)}{\chi_\mathrm{S}}\,
  a^2\rho(\chi)\;,
\label{eq:48}
\end{equation}
where Poisson's equation was used again after replacing the Laplacian with respect to perpendicular coordinates by the full Laplacian. The squared scale factor $a^2$ takes into account that we are using comoving angular-diameter distances. Expression (\ref{eq:48}) shows that the convergence is a geometrically weighted line-of-sight integral over the mass density $\rho$. It is important to note, however, that $\rho$ is the \emph{fluctuation} of the mass density about its cosmological mean value $\bar\rho$ and not the entire mass density. This is because the bending of light by the mean mass density is already incorporated into the distance measure $\chi$. The convergence $\kappa$ from (\ref{eq:48}) thus describes the lensing effects of matter inhomogeneities in an otherwise homogeneous mean universe.

Introducing the dimension-less \emph{density contrast} $\delta$ as the density fluctuation about the mean relative to the mean, the density $\rho$ to be inserted into (\ref{eq:48}) is $\rho = \bar\rho\delta$. In terms of conventional cosmological parameters, the mean matter density is
\begin{equation}
  \bar\rho = \frac{3H_0^2}{8\pi G}\Omega_\mathrm{m0}a^{-3} =
  \bar\rho_0\,a^{-3}\;,
\label{eq:49}
\end{equation}
where $H_0$ is the Hubble constant quantifying the present expansion rate of the universe, and $\Omega_\mathrm{m0}$ is the dimension-less matter-density parameter. Inserting (\ref{eq:49}) into (\ref{eq:48}) gives the expression
\begin{equation}
  \kappa = \frac{3}{2}\frac{H_0^2}{c^2}\Omega_\mathrm{m0}
  \int_0^{\chi_\mathrm{S}}\mathrm{d}\chi\,
  \frac{\chi(\chi_\mathrm{S}-\chi)}{\chi_\mathrm{S}}\,
  \frac{\delta(\chi)}{a}
\label{eq:50}
\end{equation}
for the convergence $\kappa$ of an extended lens. It is called \emph{effective convergence} because it corresponds to the convergence of a thin lens whose effects are equivalent to those caused by the actual extended matter distribution.

\section{Cosmological weak gravitational lensing}

\subsection{Limber's approximation}

The lensing effects by the cosmic large-scale structures along one particular line-of-sight cannot be predicted because the actual matter distribution in any direction is unknown. What can be predicted, however, is the degree to which lensing quantities such as the lensing potential, the deflection angle, the convergence and the shear are correlated with each other. To give just one example, this means that, if some image distortion by lensing is measured in one particular direction, the image distortion measured in a nearby direction should be similar. The smaller the angle is chosen between the two directions, the more similar the distortions are expected to be, and if the angle becomes large, the distortions should become independent.

This expected behaviour is quantified by an \emph{angular correlation function}. Let $x(\vec\theta\,)$ be a quantity measured on the sky, its angular correlation functions is
\begin{equation}
  \xi_x(\varphi) := \left\langle
    x(\vec\theta\,)x(\vec\theta+\vec\varphi\,)
  \right\rangle\;.
\label{eq:51}
\end{equation}
The average indicated by the angular brackets is quite involved: It combines an average over all positions $\vec\theta$ with an average over all orientations of the separation vector $\vec\varphi$ on the sky. The concept behind assuming that the angular correlation function depends only on the absolute value of $\varphi$ but not on its orientation is the statistical isotropy of the cosmic large-scale structures: On average, these structures should not identify an orientation on the sky.

In many applications, it is convenient to Fourier-transform the correlation function $\xi(\varphi)$ to obtain the so-called angular \emph{power spectrum},
\begin{equation}
  C(l) = \int\mathrm{d}^2\varphi\,\xi(\varphi)\,
  \mathrm{e}^{-\mathrm{i}\vec l\cdot\vec\varphi}\;.
\label{eq:52}
\end{equation}
Here, $\vec l$ is the two-dimensional wave vector conjugate to the angular separation $\vec\varphi$.

Quite often, the weak-lensing power spectra are calculated by means of Limber's approximation. It asserts that, if the quantity $x(\vec\theta\,)$ defined in two dimensions is a projection
\begin{equation}
  x(\vec\theta\,) = \int_0^{\chi_\mathrm{S}}
  \mathrm{d}\chi\,w(\chi)\,y(\chi\vec\theta,\chi)
\label{eq:52a}
\end{equation}
of a quantity $y(\vec r)$ defined in three dimensions with a weight function $w(\chi)$, then the angular power spectrum of $x$ is given by
\begin{equation}
  C_x(l) = \int_0^{\chi_\mathrm{S}}
  \mathrm{d}\chi\,\frac{w^2(\chi)}{\chi^2}\,P_y\left(\frac{l}{\chi}\right)\;,
\label{eq:53}
\end{equation}
where $P_y(k)$ is the power spectrum of $y$, taken at the three-dimensional wave number $k = l/\chi$. The condition for Limber's approximation to be applicable is that $y$ must vary on length scales much smaller than the typical length scale of the weight function $w$.

\subsection{Convergence power spectrum}

Equation (\ref{eq:50}) shows that the effective convergence is a projection of the density contrast $\delta$ with the weight function
\begin{equation}
  w(\chi) = \frac{3}{2}\frac{H_0^2}{c^2}\Omega_\mathrm{m0}
  \frac{\chi(\chi_\mathrm{S}-\chi)}{a\chi_\mathrm{S}}\;.
\label{eq:54}
\end{equation}
The power spectrum of the convergence $C_\kappa(l)$ is thus determined by a weighted line-of-sight integral over the power spectrum $P_\delta(k)$ of the density contrast.

To further clarify the physical interpretation of cosmological weak gravitational lensing, we write the power spectrum $P_\delta$ in a different way, motivated as follows. As long as the density contrast is small, $\delta\ll1$, it can be shown to increase with time in proportion to a time-dependent linear growth factor $D_+(a)$. Being quadratic in the density contrast, the power spectrum thus grows like $P_\delta\propto D_+^2$ on sufficiently large scales whose wave number $k$ is sufficiently small. On such scales, the power spectrum $P_\delta$ keeps its initial shape. On small scales, this shape is changed by non-linear effects on the evolution of the density contrast. Notwithstanding this non-linear complication, we write the power spectrum $P_\delta$ as a slowly varying shape function $\mathcal{P}$ times an amplitude. Conventionally, this amplitude is called $\sigma_8^2$ and set at the present epoch. In terms of the power spectrum linearly extrapolated to the present time, $\sigma_8$ is defined by
\begin{equation}
  \sigma_8^2 = \int_0^\infty\frac{k^2\mathrm{d}k}{2\pi^2}\,P_\delta(k)W_8^2(k)\;,
\label{eq:554a}
\end{equation} 
where $W_8(k)$ is a filter function suppressing all modes smaller than $8\,h^{-1}\,\mathrm{Mpc}$. Setting the linear growth factor $D_+$ to unity at present, the density-fluctuation power spectrum can then be written as
\begin{equation}
  P_\delta(k) = \sigma_8^2\,D_+^2(a)\,\mathcal{P}(k)\;.
\label{eq:55}
\end{equation}
With this expression for $P_\delta(k)$, the convergence power spectrum is given by
\begin{equation}
  C_\kappa(l) = \frac{9}{4}\left(\frac{H_0}{c}\right)^4
  \Omega_\mathrm{m0}^2\sigma_8^2
  \int_0^{\chi_\mathrm{S}}\mathrm{d}\chi\left[
    \frac{D_+(a)}{a}
    \frac{\chi(\chi_\mathrm{S}-\chi)}{\chi_\mathrm{S}}
  \right]^2\mathcal{P}\left(\frac{l}{\chi}\right)\;.
\label{eq:56}
\end{equation}

This equation is cosmologically very important. First, the convergence power spectrum $C_\kappa(l)$ can be measured in a way to be described shortly. Second, the \emph{shape} of the convergence power spectrum depends on the shape $\mathcal{P}$ of the power spectrum $P_\delta$ for the density contrast, which can thus be inferred from measurements as well. Third, the \emph{amplitude} of the convergence power spectrum is directly proportional to the squared matter-density parameter times the amplitude of the power spectrum, $\Omega_\mathrm{m0}^2\sigma_8^2$. Fourth, the weight function appearing in square brackets under the integral in (\ref{eq:56}) contains the linear growth factor $D_+(a)$ of the density contrast, which opens a way towards probing the evolution of cosmic structures with time.

\begin{figure}[ht]
  \centerline{\includegraphics[width=\hsize]{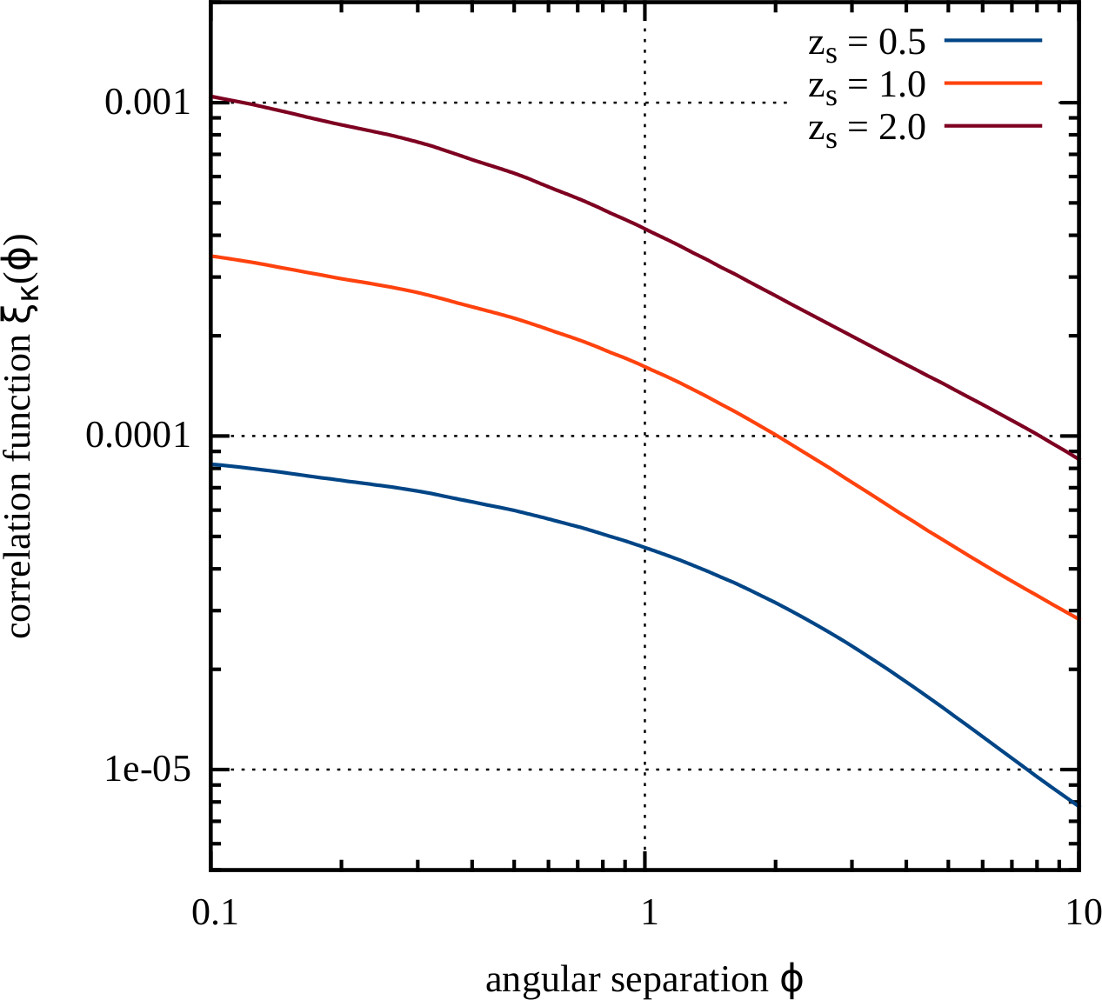}}
\caption{The angular correlation function of the effective convergence is shown for three source redshifts, as a function of the angular separation.}
\label{fig:03}
\end{figure}

The prefactor $\Omega_\mathrm{m0}^2\sigma_8^2$ is characteristic and has an intuitive meaning as well. The density parameter $\Omega_\mathrm{m0}$ quantifies the mean matter density in the universe, while the parameter $\sigma_8$ quantifies how strongly this matter is clumped. Gravitational lensing alone cannot distinguish between a low density of strongly clumped matter and a high density of weakly clumped matter.

\subsection{Measurement principle}

The convergence $\kappa$ alone can only be accessed through the magnification, as (\ref{eq:36}) shows. Measuring the magnification is possible, but difficult. Routinely, cosmological weak lensing is quantified by the shear $\gamma$. A simple consideration shows however that shear and convergence have identical power spectra. To see this, we transform the defining equations for $\kappa$ and $\gamma$ into Fourier space, where they read
\begin{equation}
  2\hat\kappa = -l^2\hat\psi\;,\quad
  2\hat\gamma_1 = -\left(l_1^2-l_2^2\right)\hat\psi\;,\quad
  \hat\gamma_2 = -l_1l_2\hat\psi\;,
\label{eq:57}
\end{equation}
with the hats denoting the Fourier transform. Now,
\begin{equation}
  4\left\vert\hat\gamma\right\vert^2 =
  \left[\left(l_1^2-l_2^2\right)^2+4l_1^2l_2^2\right]
  \left\vert\hat\psi\right\vert^2 =
  \left(l_1^2+l_2^2\right)^2\left\vert\hat\psi\right\vert^2 =
  4\left\vert\hat\kappa\right\vert^2\;,
\label{eq:58}
\end{equation}
showing that the shear power spectrum is also given by (\ref{eq:56}), $C_\gamma=C_\kappa$.

Thus, the power spectrum or, equivalently, the correlation function of the weak distortions of distant galaxy images imprinted by cosmological weak lensing needs to be measured. This can be achieved as described in the subsection ``Measurement process'' above.

\subsection{Gravitational lensing of the Cosmic Microwave Background}

The most distant source of electromagnetic radiation we can observe in our Universe is the so-called Cosmic Microwave Background (CMB). It was emitted when the cosmic radiation temperature fell to approximately $3000\,\mathrm{K}$ and allowed the cosmic plasma to recombine for the first time in cosmic history, approximately $400,000$ years after the Big Bang. As neutral atoms were formed, the free charges disappeared, and the cosmic matter became essentially transparent. From then on, the electromagnetic heat radiation left over from the Big Bang could propagate freely through the Universe.

The CMB is cosmologically highly important because, prior to its emission, energy-density fluctuations were imprinted on it by the cosmic structures that had already begun forming. We can observe these primordial energy-density fluctuations as tiny fluctuations of the CMB temperature around its mean. The mean CMB temperature today was measured by the COBE satellite to be $2.726\,\mathrm{K}$, and the relative temperature fluctuations are of order $10^{-5}$. Precise measurements of these fluctuations by the COBE, WMAP and Planck satellites as well as by several balloon-borne and ground-based experiments have turned into the major information source on the physical state of the early Universe.

However, with the CMB photons having been released about $400,000$ years after the Big Bang, they had to travel through the Universe for almost 14 billion years before they reached our detectors. Doing so, they passed the growing network of cosmic structures and experienced their weak gravitational lensing effects. The photons of the CMB could not propagate along straight lines (or unperturbed null geodesics), but along weakly perturbed paths. What we observe is thus not the CMB itself, but an image of it slightly distorted by gravitational lensing \latexonly{\cite{LE06.1}}.

Gravitational lensing of the CMB is weak and confined to angular scales of approximately 10 arc minutes and less. It can be thought of as a diffusion process slightly blurring the CMB on such angular scales. Fortunately, however, the effects of gravitational lensing on the CMB can be statistically quantified from the data itself, and thus be removed.

The essential reason for this correction to be possible is best seen thinking of a Fourier decomposition of the temperature fluctuation on the CMB. From the origin of these temperature fluctuations, their Fourier modes are independent, i.e.\ there is intrinsically no correlation between any two CMB temperature-fluctuation Fourier modes with different wave vectors. Through its focusing effect, however, gravitational lensing slightly changes the wave numbers of lensed Fourier modes. Gravitational lensing thus mixes temperature-fluctuation Fourier modes of different wave vectors and thus causes them to be correlated. In the data of the Planck satellite, the weak cosmological lensing effects on the CMB could be measured directly for the first time \latexonly{\cite{AD15.1}}. The amplitude and the shape of the lensing correlation function inferred from the CMB provide additional strong confirmation for the cosmological standard model.

\section{Weak gravitational lensing by galaxies and galaxy clusters}

\subsection{Weak lensing of galaxies by galaxies}

Less distant galaxies can act as weak gravitational lenses on more distant galaxies. Through their gravitational shear, the lensing galaxies imprint a weak tangential distortion pattern on the images of the lensed galaxies in their close neighbourhood \latexonly{\cite{BR96.1}}. This weak shear signal is superposed on the intrinsic ellipticities and irregularities of the background-galaxy images and thus requires statistical techniques for its identification and extraction. (See \latexonly{\cite{TR10.1}} for a review on strong lensing by galaxies.) Foreground and background galaxies can tentatively be separated according to their apparent brightness. Galaxy-galaxy lensing, as the effect is called, can constrain the potential depth and size of the dark-matter haloes inhabited by the lensing galaxies. First measurements in 1996 found potential depths and halo radii typical for massive galaxies \latexonly{\cite{BR96.1, DE96.1}}.

The statistical signal extraction can be improved by a thorough maximum-likelihood analysis, taking the distance distributions of lensing foreground and lensed background galaxies into account \latexonly{\cite{SC97.1}}. The galaxy-galaxy lensing signal of galaxies embedded in galaxy clusters could also be detected; it revealed that cluster galaxies are smaller than galaxies outside clusters, with their radius shrinking with the density of the environment \latexonly{\cite{NA97.2, GE98.1, GE99.1, NA02.1}}. Galaxy-galaxy lensing also reveals that the dark-matter haloes around the lensing galaxies are flattened \latexonly{\cite{HO04.1}}.

Combinations of weak gravitational lensing of galaxies by galaxies with measurements of optical brightness and stellar velocities in galaxies have been used to infer that the dark-matter distribution surrounding galaxies falls off rather steeply towards the outskirts of the dark galaxy haloes. Measurements of galaxy-galaxy lensing in wide-field surveys such as the Sloan Digital Sky Survey, the Las Campanas Redshift Survey or the Canada-France-Hawaii Legacy Survey confirm that bright galaxies have projected density profiles falling with radius $r$ approximately like $r^{-1}$, are characterised by stellar orbital velocities of approximately $v_\mathrm{c}=(150-240)\,\mathrm{km\,s^{-1}}$, are hosted by dark-matter haloes of typically $(2.7-11)\times10^{11}\,h^{-1}\,M_\odot$ and have mass-to-light ratios between $M/L\simeq120\,h\,M_\odot/L_\odot$ in blue and $\simeq170\,h\,M_\odot/L_\odot$ in red light \latexonly{\cite{FI00.1, SM01.1, WI01.1, MA06.1, PA07.1, GA07.1, MA08.1}}. 

The overall dark-matter halo masses range between $(5-10)\times10^{11}\,h^{-1}\,M_\odot$, the mass-to-light ratio increases gently with luminosity and the mass-to-light ratios of bright spiral galaxies are approximately half those of elliptical galaxies \latexonly{\cite{GU02.1,SE02.1}}. The masses quoted in this context are typically the total (virial) masses of haloes with parameterised density profiles adapted to the measured lensing signal.

Weak gravitational lensing by galaxies has also been used to study how dark matter and galaxies are correlated on large scales. The correlation function could be measured to separations reaching $10\,h^{-1}\,\mathrm{Mpc}$. It shows a typical correlation length of $r_0\simeq(5.4\pm0.7)\,h^{-1}\,\mathrm{Mpc}$ and falls of with distance with a approximate power law with exponent $1.79\pm0.05$. The density contrast in the galaxy distribution may either follow the density contrast in the dark matter, or increase slightly with scale. These results are generally in good agreement with theoretical expectations, except that the mass-to-light ratio found in simulations is typically somewhat too high. Satellite galaxies orbiting the lensing galaxies could be physically aligned with their hosts by the gravitational tidal field and thus mimic a weak galaxy-galaxy lensing signal. On the relevant scales, this possible contamination is probably less than $15\,\%$ \latexonly{\cite{HO01.1, HO02.1, YA03.1, HI04.1, SH04.1, WE04.1}}.

The availability of huge surveys with sufficient depth and image quality for weak-lensing studies has opened new applications also for galaxy-galaxy lensing. Exciting examples of more detailed studies enabled this way are the measurement of mean masses of dark-matter haloes hosting active galactic nuclei (AGN), which found that radio-loud AGN reside in host haloes which are typically $\approx20$ times more massive than for radio-quiet AGN. Moreover, the combination of galaxy-galaxy lensing with galaxy correlations has been used to specify the mean mass-to-light ratio of the galaxies and to break degeneracies between cosmological parameters \latexonly{\cite{MA09.2, CA09.1, SI13.1, MO15.1}}.

\subsection{Cluster masses and mass-to-light ratios}

As described above, galaxy clusters imprint a coherent weak distortion pattern onto the many faint and distant galaxies in their background. Since those distant galaxies reach number densities of $\simeq40$ per square arc minute in typical images taken with large ground-based telescopes, typical galaxy clusters thus cover of order $10^3$ background galaxies.

As shown above, shear and convergence are both related through the scalar lensing potential. Knowing the shear thus allows the scaled surface-mass density to be reconstructed. Cluster convergence maps can be obtained by convolving the measured shear signal with a simple kernel, opening the way to systematic, parameter-free, two-dimensional cluster studies. An immediate application of this technique to the cluster MS~1224 revealed a surprisingly high mass-to-light ratio of $\simeq800\,h\,M_\odot/L_\odot$, about four times the typical cluster value \latexonly{\cite{KA93.1, FA94.1}}

Weaknesses in this convolution algorithm mainly due to the non-locality of the convolution were identified and could be removed \latexonly{\cite{SE95.1, SE96.1}}. Another, purely local cluster reconstruction method has been proposed based on an entropy-regularised maximum-likelihood approach. This method allows the straightforward extension of the reconstruction algorithm to include all observable quantities provided by galaxy clusters \latexonly{\cite{BA96.3, SE98.1, BR05.2, BR06.1, CA06.1, ME09.2}}.

These inversion techniques for the matter distribution in galaxy-cluster lenses have by now been applied to numerous objects \latexonly{\cite{LI14.1, AP14.1, ME16.1}}. For most of them, the mass-to-light ratios turned out to be $M/L\simeq(250-300)\,h\,M_\odot/L_\odot$ in blue and $M/L\simeq(150-200)\,h\,M_\odot/L_\odot$ in red light, respectively \latexonly{\cite[see][for some examples]{CL98.1, HO02.2, GA04.1}}. Similar to weak lensing by galaxies, the masses quoted here are typically derived from parameterised density profiles adapted to the measured lensing signal and integrated to a fixed radius of order $1\,h^{-1}\,\mathrm{Mpc}$. Statistically combining the weak-lensing signal of galaxy groups, the mass range in which mass-to-light ratios can be probed could be extended towards lower masses. For galaxy groups with masses around $(10^{13}-10^{14})\,M_\odot$, values of $M/L\simeq180\,h\,M_\odot/L_\odot$ in blue and $M/L\simeq250\,h\,M_\odot/L_\odot$ in red spectral ranges have been obtained \latexonly{\cite{PA05.1, LI09.1}}.

\begin{figure}[ht]
  \centerline{\includegraphics[width=\hsize]{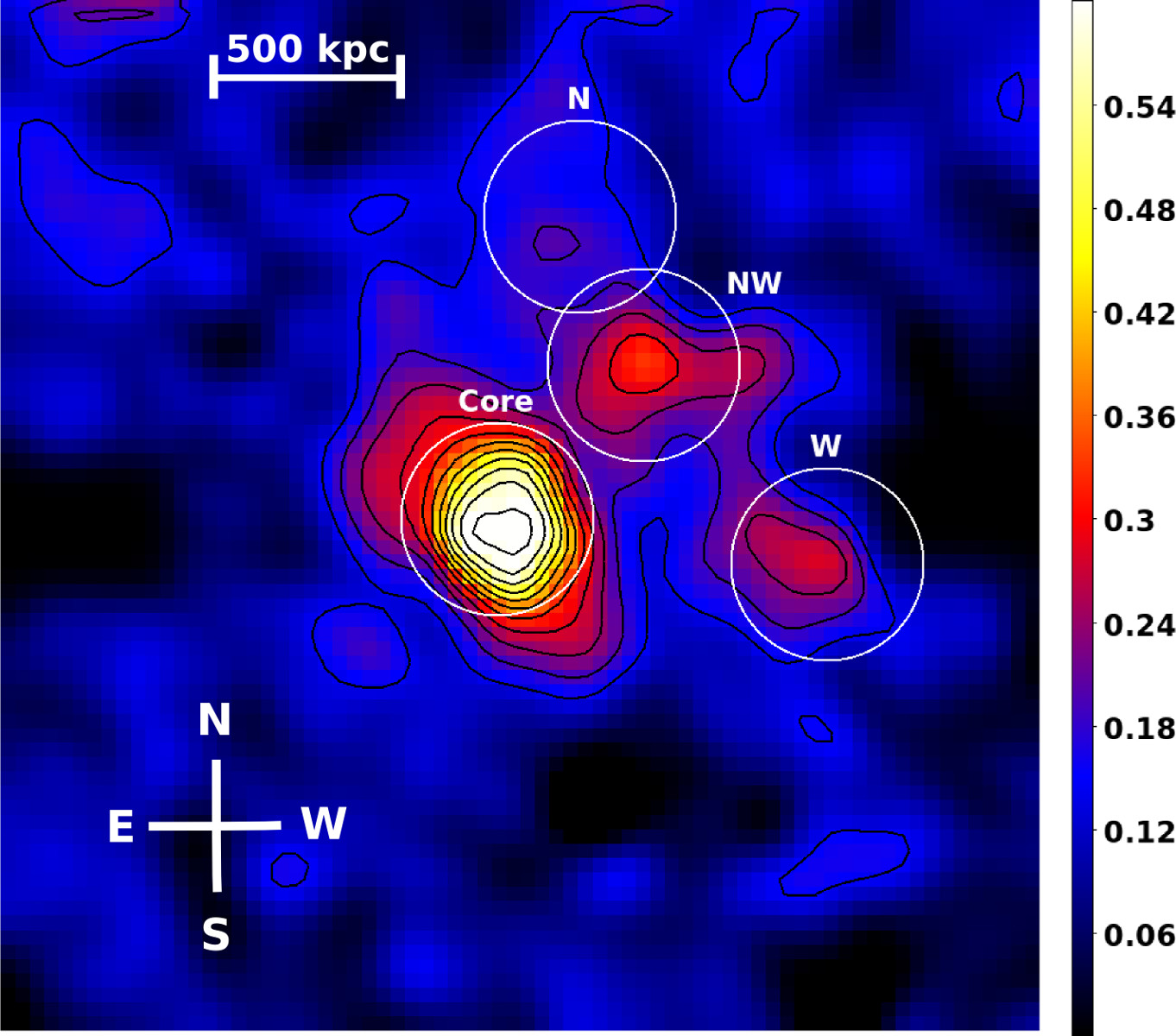}}
\caption{Mass map of the galaxy cluster Abell~2744 (sometimes called ``Pandora's box''), reconstructed from the weak-lensing signal \cite[reprinted with permission from][Fig.~3]{ME11.2}. At least two, maybe three galaxy clusters have collided here. The colour encodes the convergence $\kappa$.}
\label{fig:04}
\end{figure}

By a similar statistical analysis of gravitational lensing together with the optical light distribution, the mass-to-light ratio of the central brightest galaxies in galaxy clusters was found to be $M/L\simeq360\,h\,M_\odot/L_\odot$ \latexonly{\cite{SH09.1, SH09.2}}. Mass and light generally appear well correlated in weakly lensing clusters. Contradicting claims from individual objects could not be confirmed \latexonly{\cite{GR02.1, HE08.2}}. A very peculiar case is the galaxy cluster Abell~2744, whose galaxies and dark matter are displaced from the X-ray emission \latexonly{\cite{ME11.2}}, see Fig.~\ref{fig:04}.

\subsection{Lensing and X-ray emission}

Comparing the surface-brightness distribution of the X-rays emitted by the hot intracluster medium with the surface-density contours obtained from weak lensing, interesting phenomena are uncovered. While the X-ray surface brightness follows the matter density in many clusters (see \latexonly{\cite{GI99.1, CL00.1, HO00.1, CL02.1}} for examples), instructive deviations have been discovered.

Generally, deviations between the morphologies of the surface-mass density and the X-ray surface brightness are attributed to dynamical processes going on in galaxy clusters which are merging with other clusters or are otherwise out of equilibrium. In merging clusters, the X-ray gas is typically found lagging behind the dark matter, as expected for hot gas embedded into collision-less dark-matter halos \latexonly{\cite{MA02.2, CL04.1, MA04.3, JE05.1}}.

A particularly interesting case is the cluster 1E~0657$-$558, called the bullet cluster, whose X-ray emission appears in between two galaxy concentrations and dark-matter distributions recovered from weak lensing \latexonly{\cite{CL04.2, CL06.1, CL07.1}}. The morphology of the bullet cluster suggests that two clusters have lost their gas by friction while passing each other in the course of a merger. Other clusters showing similar morphology have since been found \latexonly{\cite{BR08.4}}.

If the hypothetical dark-matter particles interacted with each other, such a separation between gas and dark matter would be suppressed. Thus, from gas lagging behind the dark matter in merging clusters, and from small dark-matter core radii, limits could be obtained for the self-interaction cross section of the dark-matter particles, typically finding values $\lesssim(0.1-1)\,\mathrm{cm^2\,g^{-1}}$, comparable to values derived from strong gravitational lensing by clusters \latexonly{\cite{ME01.1, AR02.1, NA02.2, MA04.3, RA08.1}}.

Apart from gravitational lensing, masses of galaxy clusters can be estimated from the X-ray emission and the kinematics of the cluster galaxies. Although different mass estimates agree well in some clusters (e.g.~\latexonly{\cite{CL02.1, IR02.1, JI04.1, JE05.1, MA05.1, HO07.1, ZH08.1, LI14.1, LI14.2, AP16.1}}), substantial discrepancies are often found and interpreted as signalling dynamical processes in unrelaxed cluster cores or systematics in the data interpretation \latexonly{\cite{LE99.1, AT02.1, HO02.3, PR05.1, GI07.1, GR14.1, HO15.1, UM15.1}}. Signs of dynamical activity are often seen in massive galaxy clusters, while less massive, cooler clusters seem to be closer to equilibrium \latexonly{\cite{DA02.2, CY04.1, MA13.4, SM16.1}}.

Even though galaxy clusters are the population of cosmic objects forming last in cosmic history, spectacular examples of massive, distant clusters have been found. The weak-lensing signals of many distant clusters have been measured, typically confirming the presence of well-developed, massive and compact clusters at that epoch \latexonly{\cite{LU97.1, CL98.1, GI99.1, LO05.1, MA05.1, JE09.1}}, but also frequently indicating violent dynamical activity in cluster cores \latexonly{\cite{HO00.1, HO02.3, JE05.1, JE14.1}}.

\subsection{Cluster density profiles}

\begin{figure}[ht]
  \centerline{\includegraphics[width=\hsize]{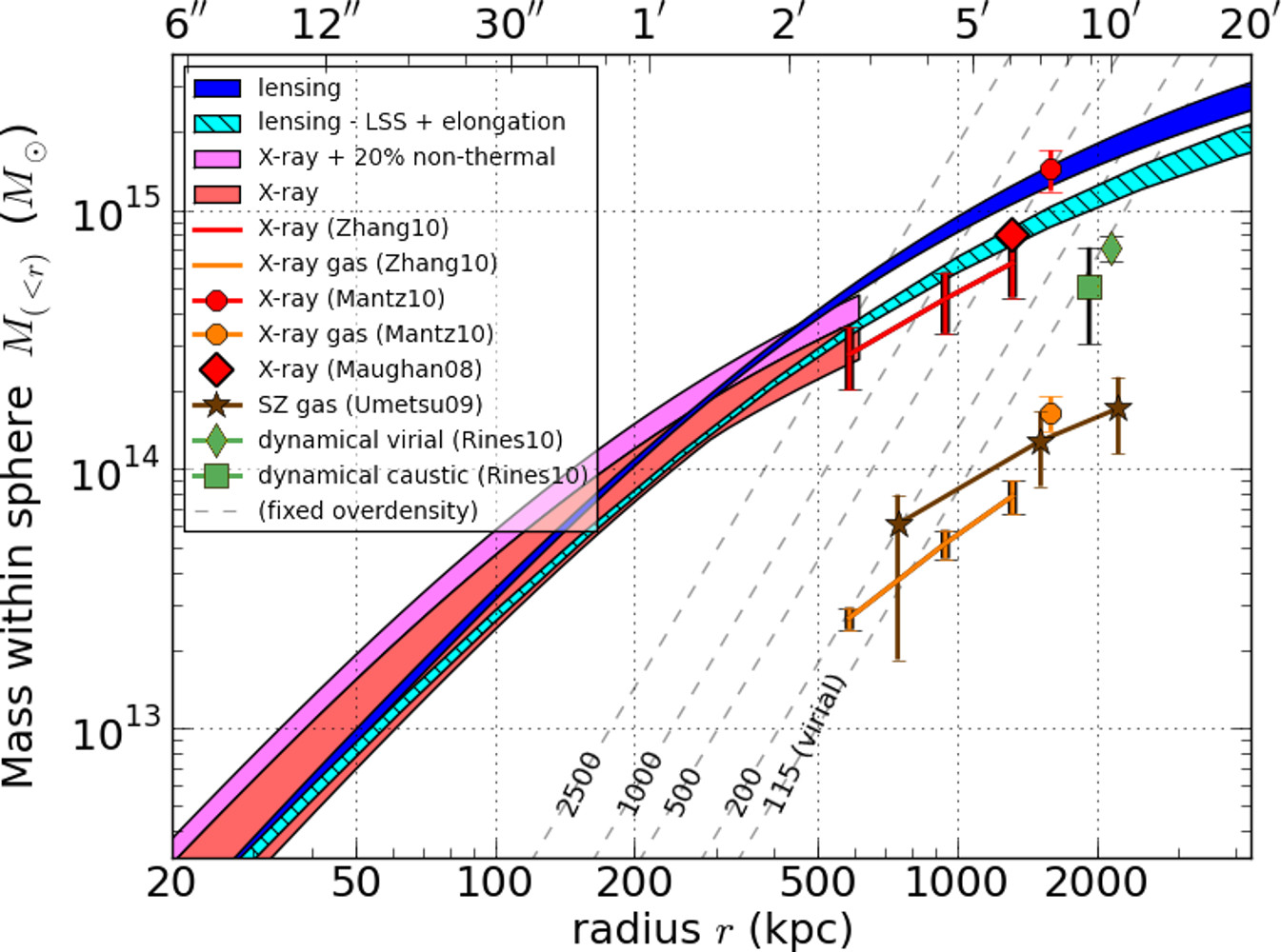}}
\caption{Radial mass profiles of the galaxy cluster Abell~2261, obtained from gravitational lensing and from different types of data \cite[reprinted with permission from][Fig.~13]{CO12.1}.}
\end{figure}

The radial matter-density profile in galaxy clusters is fundamentally important for cosmology because numerical simulations routinely show that a near-universal density profile is expected,
\begin{equation}
  \rho(x) = \frac{\rho_\mathrm{s}}{x(1+x)^2}\;,\quad
  x := \frac{r}{r_\mathrm{s}}\;.
\label{eq:58a}
\end{equation} 
It is called \emph{NFW profile} after Navarro, Frenk and White who first described it \latexonly{\cite{NA96.1, NA97.1}}. While first studies found that cluster mass-density profiles could be well fit by an isothermal profile (falling off in projection with radius $r$ as $r^{-1}$) or equally well by NFW and isothermal profiles \latexonly{\cite{CL00.1, CL01.1, SH01.1, AT02.1}}, the NFW profile became preferred as the data quality improved \latexonly{\cite{CL01.1, MA06.1, UM14.1}}.

In addition to its density parameter $\rho_\mathrm{s}$, the NFW profile is characterised by a scale radius $r_\mathrm{s}$. Simulations find that the ratio between the overall (virial) radius and the scale radius, called concentration, is gently decreasing with mass. While cluster profiles derived from weak gravitational lensing alone tend to have somewhat larger scale radii than expected, possibly reflecting deviations from spherical symmetry \latexonly{\cite{CL00.1, CL02.1, HO02.2, JE05.1, MA08.1}}, scale radii well in the expected range are typically derived from data of high quality analysed with sophisticated methods \latexonly{\cite{CL01.1, AR02.1, DA03.1}}. An exciting step beyond determining density profiles is the measurement of the subhalo mass function in the Coma galaxy cluster \latexonly{\cite{OK14.1}}.

The large-scale matter distribution in front of and behind galaxy clusters is projected onto them and can affect weak-lensing mass determinations. Estimates based on simulations and analytic calculations indicate that cluster mass estimates from weak gravitational lensing can be changed at the level of up to 40 per cent, and that uncertainties of the total mass estimates can be approximately doubled by projection. Cluster density profiles, however, should only be weakly affected \latexonly{\cite{ME99.1, KI01.1, HO03.1, CL04.1}}.

\subsection{Cluster detection}

Several detections of clusters with very high mass-to-light ratios have been claimed and raised the question whether cluster-sized dark-matter halos may exist which are so inefficient in producing stellar or X-ray emission that they are invisible to anything but gravitational lensing \latexonly{\cite{FA94.1, FI99.1, ER00.1, UM00.1, GR01.1, MI02.2}}. The most prominent cases discussed of potentially dark clusters so far, however, did not outlast later re-analyses of the data or analyses of new data \latexonly{\cite{GR02.1, ER03.1, HE08.2}}.

Weak gravitational lensing also provides a powerful way to detect galaxy clusters regardless of their directly observable signatures. Methods developed for this purpose use weighted integrals over circular apertures of the gravitational shear signal tangentially oriented with respect to the aperture centre. Numerical simulations show that these methods are highly efficient in finding suitably massive matter concentrations if parameters and weight functions are optimally chosen to carefully balance the completeness against the frequency of spurious detections \latexonly{\cite{SC96.1, RE99.1, WH02.1, MA05.2, PA07.2}}. Substantial samples of galaxy clusters have been routinely detected and confirmed by this technique. Their statistical analysis is likely to provide important information on the origin and evolution of non-linear cosmic structures in the near future \latexonly{\cite{KR99.1, WI01.2, BA01.1, MI02.3, SC04.2, HE05.1, WI06.1, DI07.1, GA07.2, MI07.1, MA07.3, SC07.1, DI10.1, KR10.1, MA10.1, MA11.1, SH14.1, LI15.1}}. Sufficiently numerous and well-defined cluster samples may well be the most sensitive probe into a possible time dependence of the enigmatic dark energy supposed to drive our Universe apart in an accelerated fashion \latexonly{\cite{ER00.1, UM00.1, BA02.1, WE02.1, WE03.1, MA13.3, KI15.2, RE16.1}}.

Galaxy clusters are embedded into a network of filamentary structures \latexonly{\cite{BO96.1, EB04.1, CO05.2, HA07.1}} which may be traced by direct mapping or with linear filtering techniques similar to those developed for halo detection, albeit with generally low signal-to-noise ratio \latexonly{\cite{DI05.1, ME10.1, JA12.1, MA13.2}}.

\section{Weak gravitational lensing by large-scale structures}

\subsection{Expectations and measurements}

Weak gravitational lensing by large-scale structures, or cosmological weak lensing, is a rich and rapidly developing field which is covered in detail by several dedicated reviews \latexonly{\cite{ME99.2, BA01.2, HO02.5, ME02.4, RE03.1, HO08.1, MU08.1}}. We summarise the most important aspects here and refer the interested reader to those reviews for further detail.

Early studies used analytic calculations to show that the standard deviation of image ellipticities induced by cosmological weak lensing were of order a few per cent on arc-minute angular scales. A first attempt at measuring this tiny signal placed an upper limit in agreement with theoretical expectations \latexonly{\cite{BL91.1, MI91.1, KA92.2, MO94.1}}. Since weak cosmological lensing depends sensitively on the non-linear evolution of the large-scale structures, numerical simulations are required for precisely estimating the expected amplitude of the signal and the shape of the ellipticity correlation function \latexonly{\cite{BA92.1, JA97.1, JA00.1, HA01.1, VA03.1, HI09.1}}. The cosmological potential of large weak-lensing surveys was quickly pointed out \latexonly{\cite{BE97.1, KA98.1, WA99.1}}, emphasising the possibility of measuring in particular the matter density parameter $\Omega_\mathrm{m0}$ and the amplitude $\sigma_8$ of the dark-matter power spectrum.

The first detections of cosmological weak lensing were announced around the turn of the century \latexonly{\cite{SC98.1, BA00.1, WA00.1, WI00.1, MA01.1}}. Given the enormous difficulty of the measurement and the different telescopes, cameras, and analysis techniques used, the agreement between these results and their compatibility with theoretical expectations was surprising and encouraging at the same time.

Assuming a spatially flat cosmological model, the cosmological parameters $\Omega_\mathrm{m0}$ and $\sigma_8$ were derived from measurements of the cosmic-shear correlation function. Since these parameters are degenerate, higher than two-point statistics are needed for constraining them separately \latexonly{\cite{WA01.1, WA01.2, BE02.1, KI05.1}}. Enormous effort was subsequently devoted to calibrating weak-lensing measurements, to designing optimal cosmic-shear estimators and studying their noise properties, and to shaping theoretical expectations \latexonly{\cite{ER01.1, HU01.1, CO01.1, SC02.3, SE07.2, JO08.1, HA09.1}}.

Numerous weak-lensing surveys have meanwhile been conducted, and cosmological parameters derived from them. Values obtained for the matter-fluctuation amplitude $\sigma_8$ at fixed $\Omega_\mathrm{m0} = 0.3$ tend to agree within the error bars, but the scatter is still substantial: They fall into the range $0.65-1.06$, with much of the uncertainty due to remaining systematics in the data analysis \latexonly{\cite{MA01.1, HO02.4, BA03.2, BR03.1, HA03.1, JA03.1, RH04.1, HE05.2, MA05.3, WA05.1, HO06.1, SE06.1, BE07.1, HE07.2, KI07.1, MA07.4, SC07.2, FU08.1, HA09.1}}.

\begin{figure*}[ht]
  \centerline
   {\includegraphics[width=0.49\hsize]{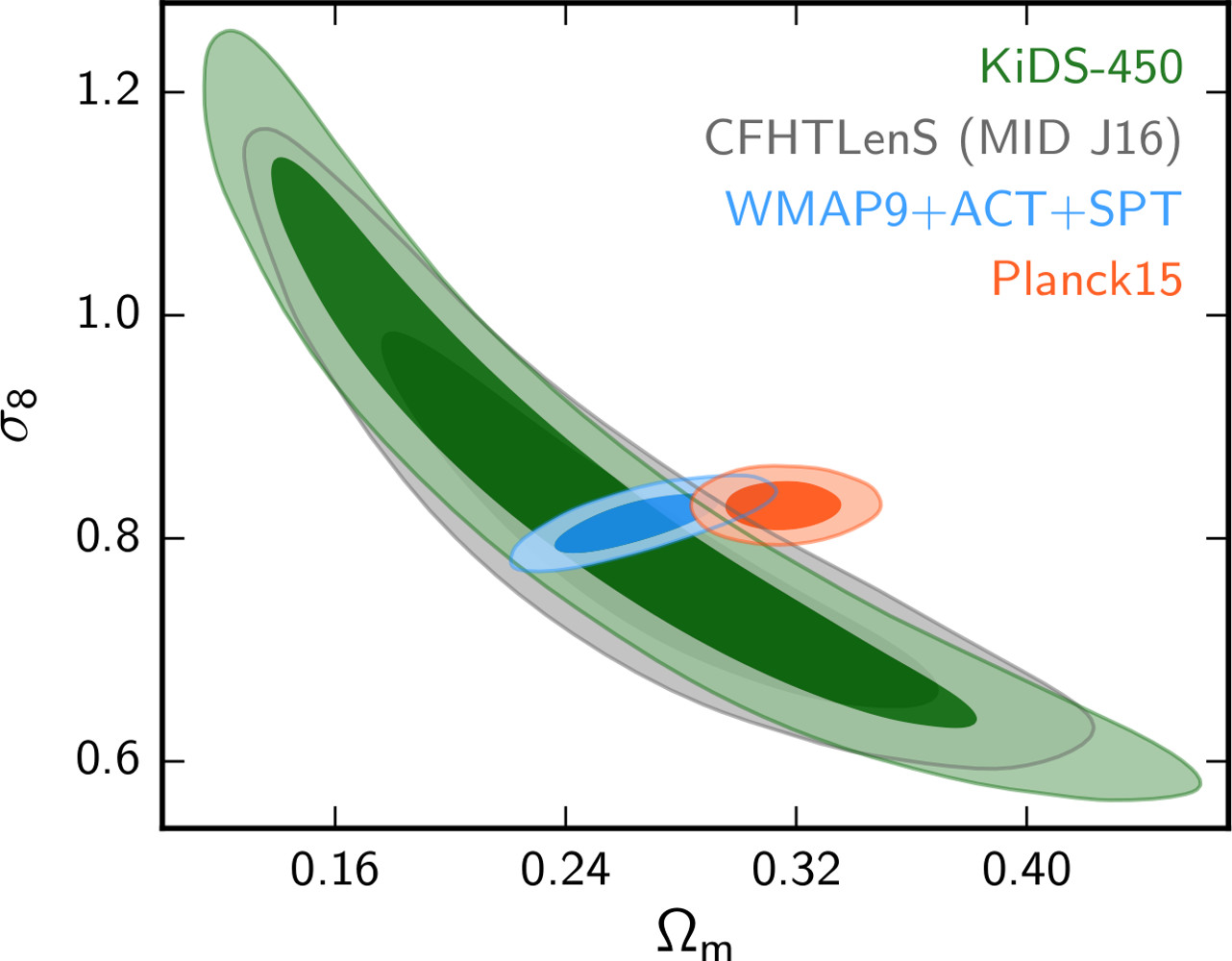}\hfill
    \includegraphics[width=0.49\hsize]{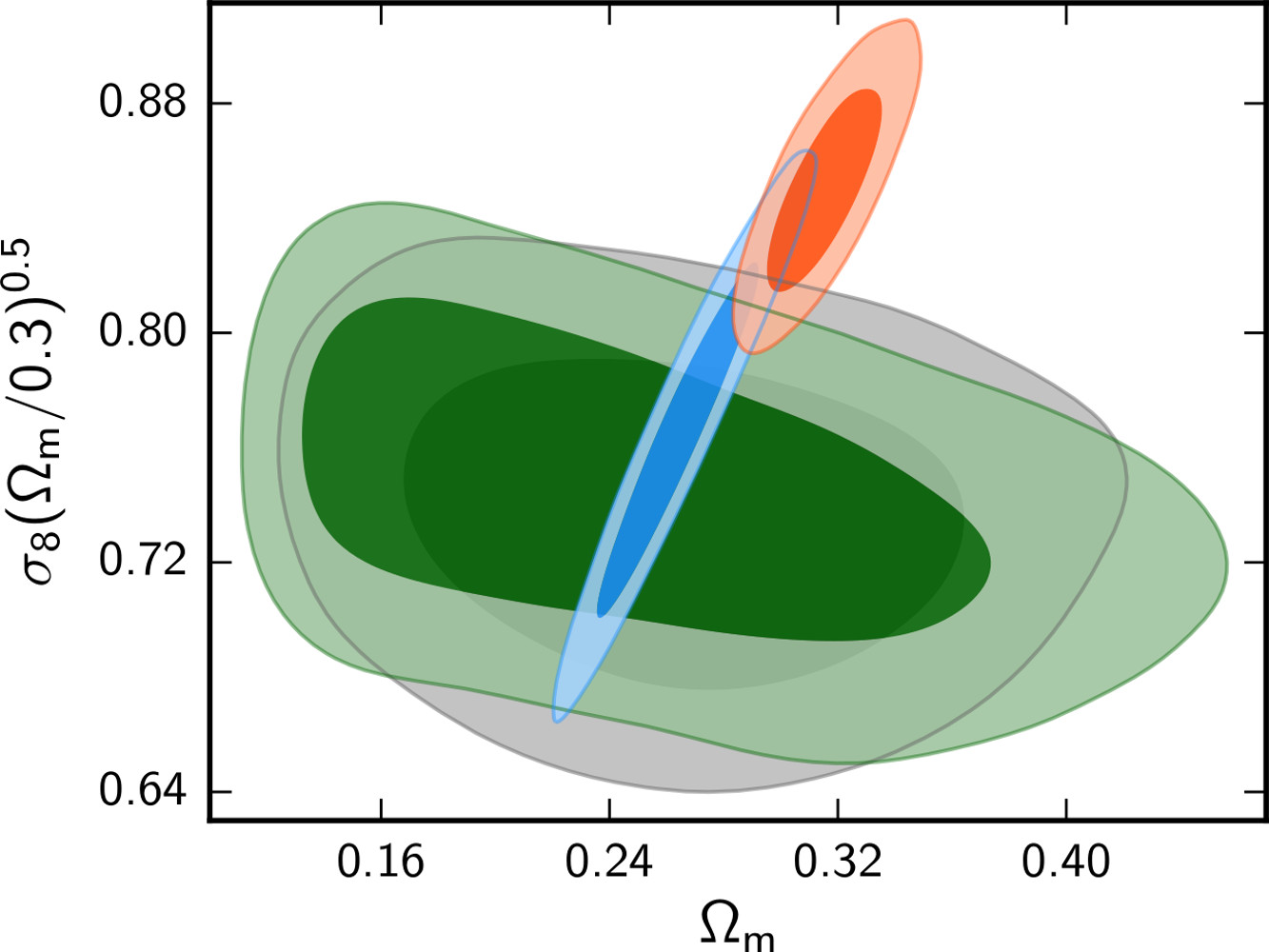}}
\caption{Constraints on two cosmological parameters, the normalisation parameter $\sigma_8$ and the matter-density parameter $\Omega_\mathrm{m0}$, obtained from cosmic-shear correlation functions measured from $\sim450$ square degrees of the KiDS survey \latexonly{\cite[reprinted with permission from][Fig.~6]{HI16.1}}. The green area in the panel shows constraints from the KiDS survey, the grey area those from the CFHTLens survey \latexonly{\cite{HE13.2}}. In the right panel, the vertical axis shows the combined parameter $\sigma_8(\Omega_\mathrm{m0}/0.3)^{0.5}$ that weak cosmological lensing is sensitive to. Blue and red areas show the independent constraints from CMB measurements prior to Planck and by Planck, respectively.}
\label{fig:06}
\end{figure*}

\begin{figure}[ht]
  \centerline{\includegraphics[width=\hsize]{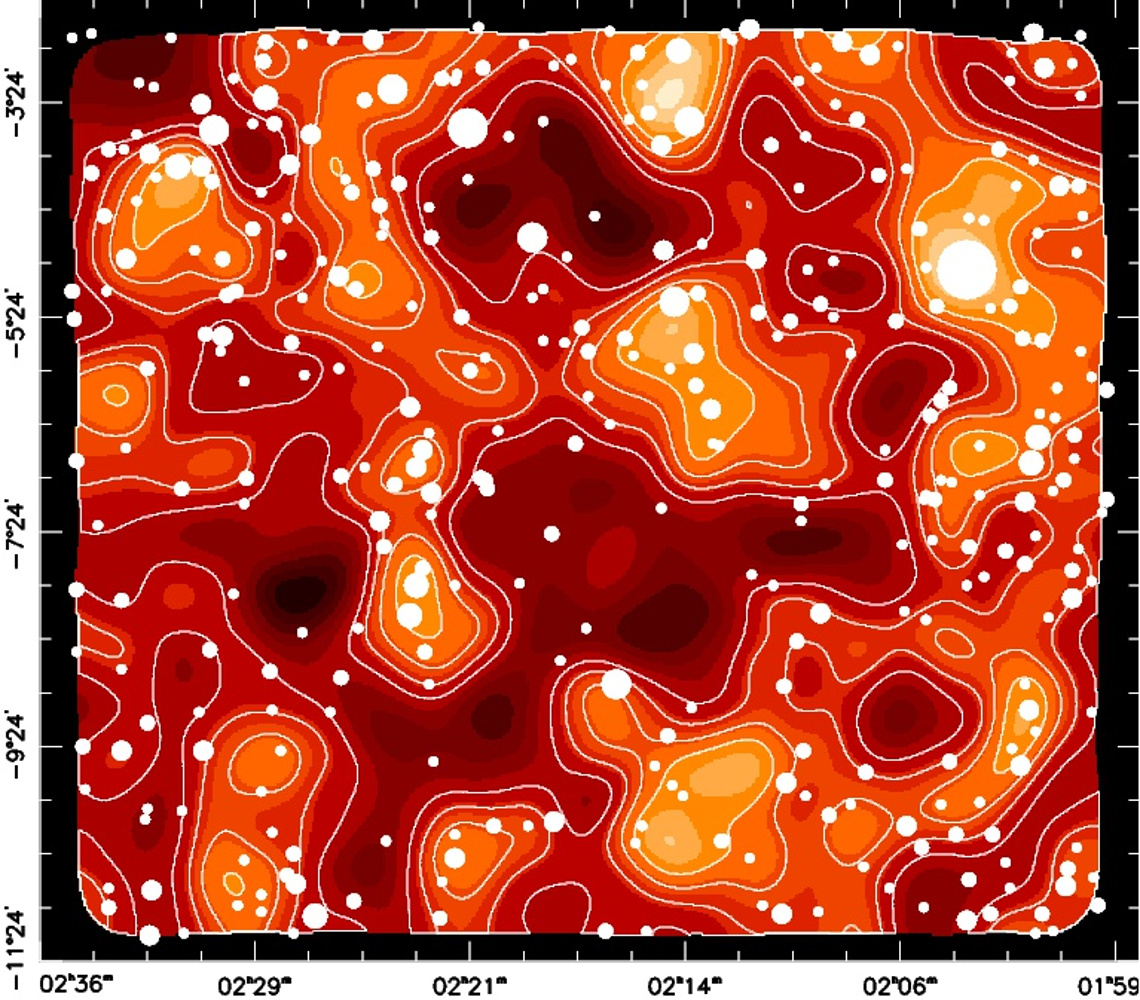}}
\caption{Map of matter distribution in one of the CFHTLens fields, reconstructed from weak gravitational lensing \cite[][Fig.~8]{WA13.1}. The white dots mark positions of local galaxy concentrations.}
\label{fig:07}
\end{figure}

Dedicated weak-lensing surveys of increasing portions of the sky have been concluded with remarkable success, are ongoing or being planned, with the solid angles covered by the surveys increasing in steps of a factor of ten. The Canada-France-Hawaii Telescope Lensing Survey (CFHTLens) has covered 154 square degrees \latexonly{\cite{HE12.1, WA13.1}}; see Figs.\ \ref{fig:06} and \ref{fig:07} for selected results. The Kilo-Degree Survey \latexonly{\cite{JO13.1}} and the Dark-Energy Survey \latexonly{\cite{DE05.1}} are overing 1500 and 5000 square degrees, respectively. The Large Synoptic Survey Telescope and the Euclid satellite are planned to cover approximately 15000 square degrees \latexonly{\cite{LA12.1, CH13.1}}. Besides mapping dark matter and determining cosmological parameters, a primary goal of these surveys is to constrain the growth of cosmological structures and thereby the nature of the dark energy. Future wide-field radio surveys will also offer exciting possibilities for cosmological applications of weak gravitational lensing \latexonly{\cite{BR15.1, BO16.1, PA16.1}}.

\subsection{Systematics}

Numerical simulations show that the Born approximation is valid to a very high degree \latexonly{\cite{DO05.1, SH06.1, HI09.1}}. As shown above, the gravitational lensing effects even by widely extended lenses can be summarised by a scalar potential. Thus, weak lensing can only cause such distortion patterns which can be described by derivatives of a scalar potential. This motivates the decomposition of distortion patterns in gradient and curl contributions, termed $E$ and $B$ modes in analogy to electrodynamics. Significant $B$ modes in the data are interpreted as remainders of undetected or incompletely removed systematics. More or less significant $B$ modes have been found in almost all weak-lensing surveys. Meticulous studies revealed that they originated from various effects, among them incomplete correction of astigmatism in the telescope optics, source clustering or finite-field effects. Once these reasons had been identified, methods were developed for removing the $B$ modes they caused \latexonly{\cite{SC02.2, HO04.2, WA05.1, KI06.1, SC07.3, FU10.1}}.

At least five important sources of systematic error plague weak-lensing measurements: astigmatism of the telescope optics, miscalibrated distortion measurements, insufficient knowledge of the non-linear density-fluctuation power spectrum, insufficient information on the distance distribution of the lensed galaxies, and intrinsic alignments of background sources. All these effects have been addressed in detail, and sophisticated techniques have been developed for removing them from the data or at least for quantifying them \latexonly{\cite{GU05.1, HE06.1, JA06.1, JI06.1, WA06.1, MA07.1, BR09.2, HI09.1, OS15.1, TR15.1}}. By now, the largest sources of systematic error are probably the distances of the background sources, the non-linear evolution of the power spectrum and the influence of baryonic matter on the dark-matter distribution.

The potentially harmful effect of intrinsic rather than lensing-induced galaxy alignments depends obviously on the depth of the survey. Deep surveys project galaxy images along light paths which are substantially longer than any large-scale structure correlation scale and thus suppress any spurious signal due to intrinsinc alignments of physically neighbouring galaxies. In shallow surveys, however, intrinsic source alignments may substantially contaminate any weak-shear signal. The degree to which intrinsic alignments affect weak-lensing measurements is still a matter of lively debate in both theory and observation, and between both disciplines \latexonly{\cite{CR00.1, HE00.1, CA01.1, CR01.1, BR02.2, JI02.1, LE02.1, LE04.1, FA09.1, BR10.1, JO11.1, MA11.2, KI12.1, CA13.1, HE13.2, JO13.2, ME13.1, GI14.1, ME14.1, KI15.1, SC15.1, SI15.1, SI15.2}}.

Likewise, possibilities for removing the signal contamination due to intrinsic alignments are being discussed extensively. They advocate using approximate photometric distance information to remove physically close pairs of source galaxies from the analysis. Applications of this technique suggest that the effects of intrinsic alignments should be near the lower end of the theoretical predictions. Foreground galaxies are additionally aligned with the large-scale structures lensing the background galaxies, thus giving rise to an indirect alignment between galaxies at different redshifts. This further effect can substantially reduce the measured shear signal, leading to a likely underestimate of the $\sigma_8$ parameter by several per cent \latexonly{\cite{KI02.2, BR03.1, HE03.1, HE04.1, HI04.2, MA06.5, HI07.1}}.

\subsection{Perspectives}

Approximately since the turn of the century, cosmology has a standard model, most of whose parameters are now known at the per-cent level or better. To a large degree, this was enabled by the precise measurements of the temperature and polarisation fluctuations in the CMB, combined with surveys of the large-scale galaxy distribution and measurements of the cosmic expansion rate by means of type-Ia supernovae. What is the role of weak gravitational lensing in this context?

Cosmological parameter constraints from the CMB alone suffer from degeneracies which can only be broken using additional information. By measuring the dark-matter density and the normalisation of its fluctuation amplitude directly, gravitational lensing adds constraints which substantially narrow the parameter ranges allowed by the CMB alone \latexonly{\cite{HU99.1, CO03.1, TE05.1}}. Also, the exploitation of higher than two-point statistics helps in breaking degeneracies in the weak-lensing parameter estimates and in analysing deviations from the primordial Gaussian statistics that develop from non-linear structure growth \latexonly{\cite{ZH03.1, TA03.1, DO04.1}}.

Perhaps the most exciting promise of weak gravitational lensing by large-scale structures derives from its potential to study the three-dimensional distribution of dark structures, originating from the distance dependence of the lensing observables. Although these observables measure the tidal field of the two-dimensional, projected matter distribution, selecting sources at multiple distances allows structures along the line-of-sight to be resolved. The distances to the source galaxies can be estimated with sufficient accuracy by photometric (rather than spectroscopic) methods. Such estimates can be used to group the source-galaxy population by distance shells and thus to extract three-dimensional information on the lensing matter distribution \latexonly{\cite{HE03.2, PE04.2, SI04.1, TA04.1, BE13.1, WA13.1, KI14.1}}.

Even poorly resolved three-dimensional information from weak gravitational lensing constrains the growth of cosmic structures along the line-of-sight from the distant and past universe until the present. Sufficiently precise measurements of weak lensing in wide fields on the sky should thus enable accurate constraints on the dynamics of the accelerated expansion of the Universe. This constitutes the strongest motivation for weak-lensing surveys on increasing areas and from space, and for getting remaining systematics under ever better control \latexonly{\cite{HU02.1, MU03.1, BE04.1, CH04.1, MA04.4, PE04.3, RE04.1, KI05.2, HE06.2, BR07.1, KI08.1, RE09.1, SE16.1}}. A further exciting perspective for gravitational-lensing research is the possibility to test the theory of gravity itself \latexonly{\cite{FE08.2, RE10.1, SI13.2, BL16.1}}.

\subsection{Cosmic magnification}

Gravitational lensing not only distorts the images of distant galaxies, but also magnifies them. As shown above to linear order, the power spectrum of the magnification is just four times the power spectrum of the gravitational shear, hence both magnification and shear contain the same amount of information. The gravitational shear, however, can much more easily be measured than the magnification because the ellipticities of distant galaxies average to zero while the intrinsic flux of any given source is generally unknown.

Currently the most promising method for detecting gravitational magnification rests on the so-called magnification bias. If a population of distant sources is observed within a certain solid angle $\delta\Omega$ on the sky where the magnification is $\mu$, fainter sources become visible there. At the same time, their number density is reduced because the solid angle is enlarged by the magnification. The net effect depends on how many more sources the magnification lifts above the sensitivity threshold of the observation. If the number of visible sources increases more than linearly with the magnification, the dilution by the enlarged solid angle is outweighed and the magnification causes more sources to become visible.

There is a class of intrinsically bright sources in the distant universe, the so-called quasi-stellar objects or QSOs, which appear more numerous on magnified areas of the sky. At the same time, the large-scale structures responsible for gravitational lensing contain more galaxies where the matter is more concentrated. This correlation of galaxies with the magnifying lenses, together with the increase of QSO number counts due to the lensing magnification, creates a measurable, apparent correlation between distant QSOs and foreground galaxies. Due to the non-linearity of the magnification in the convergence and the shear, accurate theoretical predictions for cosmic magnification are more difficult than for cosmic shear. Typical magnifications are of order 10 per cent for moderately distant sources. In addition to QSOs, certain populations of distant galaxies can also be used as sources \latexonly{\cite{BA95.3, DO97.1, MO98.1, MO98.2, GU01.1, JA02.1, BA03.3, ME03.3, TA03.2}}.

The existence of correlations between distant QSOs and foreground galaxies could quickly be established \latexonly{\cite{SE79.1, FU90.1, BA93.1, BA93.2, BA94.3, BA94.2, BA97.1}}. It was considerably more difficult to determine accurate expected amplitudes and angular scales of these correlations and finally converge on results that could be understood theoretically, including systematic effects due to extinction by intervening dust and fluctuations in the correlation between the foreground galaxies and the gravitationally-lensing large-scale structures \latexonly{\cite{RO94.1, FO96.1, FE97.1, WI98.1, NO00.1, BE01.1, CR01.2, NO01.1, GA03.2, JA03.3}}.

The weakness of the signal and the interference of a variety of confusing effects required large surveys taken in several photometric bands to unambiguously detect cosmic magnification and to extract cosmological information from it, but the information gain is expected to be substantial \latexonly{\cite{BE99.1, ME02.3, ME03.4, ME03.5, ME05.1, SC05.1, ME08.2, HI09.2, WA10.1, SC12.1, HE13.1, DU14.1, AL15.1, GI16.1}}.

\subsection{Gravitational lensing of the CMB}

The effects of gravitational lensing on the CMB are rich in detail \latexonly{\cite{LE06.1}}. Pioneering studies showed that the CMB is expected to be measurably lensed by cosmic structures in a way which resembles a diffusion process and builds up small-scale structure at the same time \latexonly{\cite{CA93.1, CA93.2, SE96.2, SE96.3, BE97.2, CA97.1, ME97.1, BE98.1, MA05.5, HA15.1}}. As mentioned above, gravitational lensing of the CMB can be identified by the coupling lensing creates between different Fourier modes of the temperature-fluctuation pattern in the CMB \latexonly{\cite{GU00.1, HU00.1, HU01.2, CH02.3, KE02.2, HI03.1, HI03.2, OK03.1, CH05.1}}.

\begin{figure}[ht]
  \centerline{\includegraphics[width=\hsize]{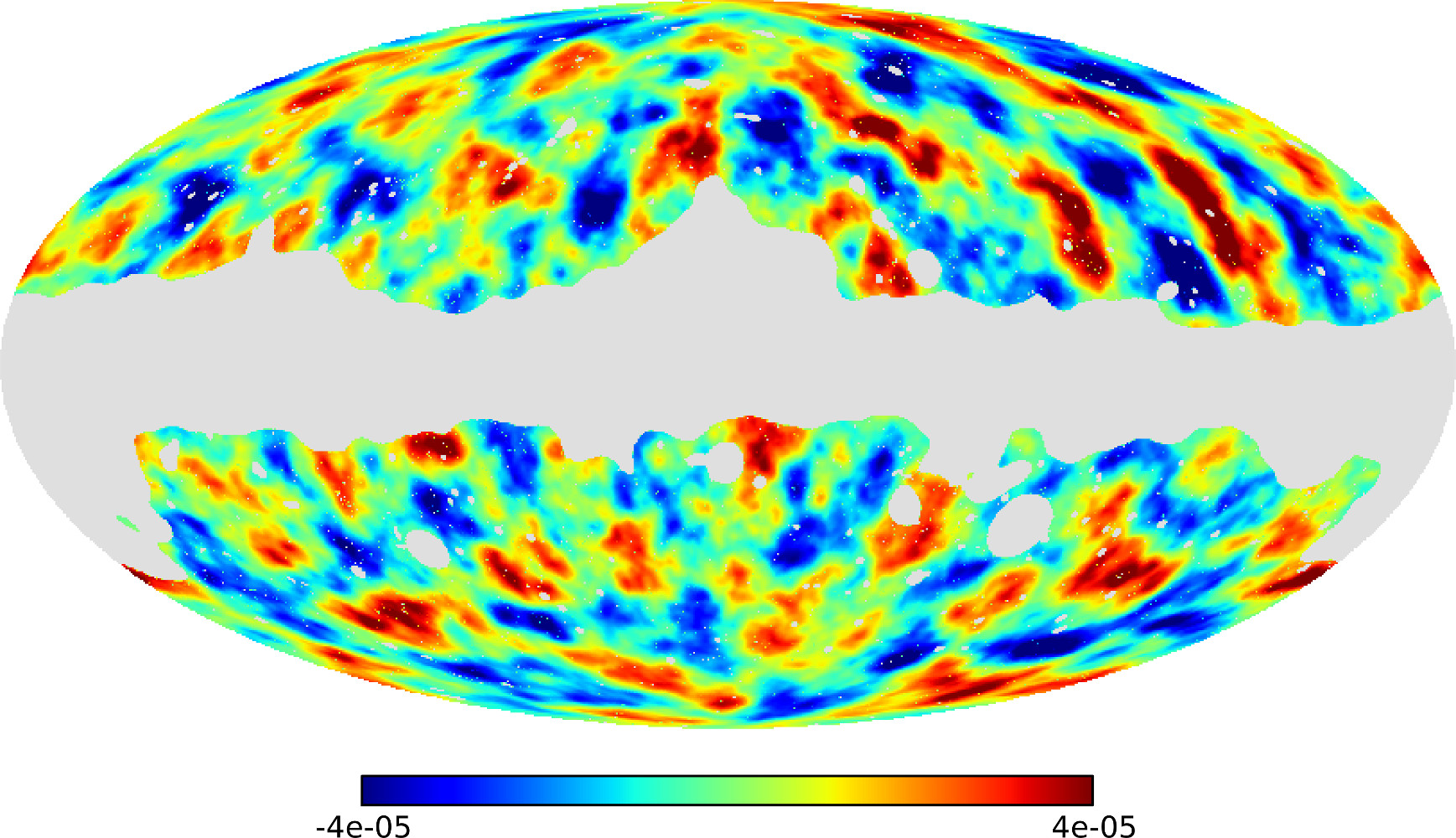}}
\caption{Sky map of the gravitational-lensing potential reconstructed from the CMB temperature fluctuations as measured by the Planck satellite \cite[reprinted with permission from][Fig.~2]{AD15.1}.}
\label{fig:08}
\end{figure}

Accurate constraints of cosmological parameters from well-resolved CMB temperature fluctuations are possible only if the weak gravitational lensing effects can be quantified and corrected. At the same time, the gravitational lensing signal contained in the CMB itself can be used to break parameter degeneracies in the purely primordial CMB data \latexonly{\cite{ST99.1, LE05.1, CA08.1, CA09.2, PE09.1}}. After indirect evidence had been found for gravitational lensing of the CMB \latexonly{\cite{SM07.2, HI08.1}}, the direct and unambiguous detection of cosmological weak lensing in the CMB data was among the most fascinating results of the Planck satellite mission \latexonly{\cite{PL14.1, AD15.1, AD15.2}}.

\section{Summary}

Gravitational lensing has two major advantages that turn it into one of the most versatile, contemporary cosmological tools: its foundation in the theory of gravity is reasonably straightforward, and it is sensitive to matter (and energy) inhomogeneities regardless of their internal physical state. Under the assumptions that gravitational lenses are weak, move slowly with respect to the cosmological rest frame, and are much smaller than cosmological length scales, the effects of gravitational lensing are entirely captured by a two-dimensional effective lensing potential. The Hessian matrix of this potential defines the local imaging properties of gravitational lenses. The Poisson equation, relating the Laplacian of the potential to the surface-mass density, allows one to infer the lensing matter distribution from observable image distortions.

Weak gravitational lensing has been applied on a broad range of scales. Weak lensing of galaxies by galaxies was used to infer mass-to-light ratios of galaxies, their sizes, and constraints on their radial density profiles. Besides constraining the density profile and the mass-to-light ratio, weak lensing by galaxy clusters has allowed us to map the spatial distribution of dark and luminous matter in clusters as well as to constrain the amount of substructure in them. In spectacular examples, it was possible to show that dark and luminous matter must have been separated in cluster collisions, yielding tight upper limits on the self-interaction cross section of the hypothetical dark-matter particles. On yet larger scales, cosmological weak lensing is now routinely being used to constrain the amount of dark matter and the amplitude of its fluctuations, and also to map the matter distribution in wide fields on the sky. Finally, it has recently become possible to measure the cosmological weak lensing effect on temperature fluctuations in the CMB and to create sky maps of what could be called the index of refraction of the entire visible universe.

\section*{Acknowledgements}

This work was supported in part by the Transregional Collaborative Research Centre TR 33 ``The Dark Universe'' of the German Science Foundation (DFG). We are grateful to Olivier Minazzoli for inviting us to write this contribution to Scholarpedia and for his patience with our very slow return. We also wish to thank the many colleagues who helped us shape our views on weak lensing with enlightening and enjoyable discussions, way too many to name them all individually -- they form a good fraction of the authors cited below! (Unfortunately not including Sir Isaac Newton.)

{\footnotesize
 \bibliographystyle{unsrtnat}
 \bibliography{main}}

\end{document}